\documentclass[twoside,twocolumn,9pt]{article}
\usepackage{comment}
\usepackage{extsizes}
\usepackage[super,sort&compress,comma]{natbib} 
\usepackage[version=3]{mhchem}
\usepackage[left=1.5cm, right=1.5cm, top=1.785cm, bottom=2.0cm]{geometry}
\usepackage{balance}
\usepackage{mathptmx}
\usepackage{sectsty}
\usepackage{stmaryrd}
\usepackage{graphicx} 
\usepackage{lastpage}
\usepackage[format=plain,justification=justified,singlelinecheck=false,font={stretch=1.125,small,sf},labelfont=bf,labelsep=space]{caption}
\usepackage{float}
\usepackage{fancyhdr}
\usepackage{fnpos}
\usepackage[english]{babel}
\usepackage{pgfplots}
\usetikzlibrary{calc}
\usepackage{dsfont}

\addto{\captionsenglish}{%
  
}
\usepackage{array}
\usepackage{droidsans}
\usepackage{charter}
\usepackage[T1]{fontenc}
\usepackage{setspace}
\usepackage[compact]{titlesec}
\usepackage{hyperref}
\usepackage{amsfonts}

\usepackage{epstopdf}

\definecolor{cream}{RGB}{222,217,201}

\newcommand{\ff}{\mathbf{f}}
\newcommand{\eeta}{\boldsymbol{\eta}}
\newcommand{\nn}{\mathbf{n}}
\renewcommand{\tt}{\mathbf{t}}

\newcommand{\rr}{\mathbf{r}}
\newcommand{\RR}{\mathbb{R}}
\renewcommand{\ss}{\mathbf{s}}
\newcommand{\uu}{\mathbf{u}}
\newcommand{\TT}{\mathbf{T}}
\newcommand{\UU}{\mathbf{U}}
\newcommand{\xx}{\mathbf{x}}
\newcommand{\yy}{\mathbf{y}}

\newcommand{\subfigimg}[3][,]{%
  \setbox1=\hbox{\includegraphics[#1]{#3}}
  \leavevmode\rlap{\usebox1}
  \rlap{\hspace*{0pt}\raisebox{\dimexpr\ht1-0\baselineskip}{\bf
  \normalsize #2}}
  \phantom{\usebox1}
}

\begin{document}

\pagestyle{fancy}
\thispagestyle{plain}
\fancypagestyle{plain}{
\renewcommand{\headrulewidth}{0pt}
}

\makeFNbottom
\makeatletter
\renewcommand\LARGE{\@setfontsize\LARGE{15pt}{17}}
\renewcommand\Large{\@setfontsize\Large{12pt}{14}}
\renewcommand\large{\@setfontsize\large{10pt}{12}}
\renewcommand\footnotesize{\@setfontsize\footnotesize{7pt}{10}}
\makeatother

\renewcommand{\thefootnote}{\fnsymbol{footnote}}
\renewcommand\footnoterule{\vspace*{1pt}%
\color{cream}\hrule width 3.5in height 0.4pt \color{black}\vspace*{5pt}} 
\setcounter{secnumdepth}{5}

\makeatletter 
\renewcommand\@biblabel[1]{#1}            
\renewcommand\@makefntext[1]%
{\noindent\makebox[0pt][r]{\@thefnmark\,}#1}
\makeatother 
\renewcommand{\figurename}{\small{Fig.}~}
\sectionfont{\sffamily\Large}
\subsectionfont{\normalsize}
\subsubsectionfont{\bf}
\setstretch{1.125} 
\setlength{\skip\footins}{0.8cm}
\setlength{\footnotesep}{0.25cm}
\setlength{\jot}{10pt}
\titlespacing*{\section}{0pt}{4pt}{4pt}
\titlespacing*{\subsection}{0pt}{15pt}{1pt}

\fancyfoot{}
\fancyfoot[LO,RE]{\vspace{-7.1pt}\includegraphics[height=9pt]{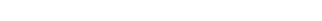}}
\fancyfoot[CO]{\vspace{-7.1pt}\hspace{13.2cm}\includegraphics{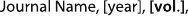}}
\fancyfoot[CE]{\vspace{-7.2pt}\hspace{-14.2cm}\includegraphics{head_foot/RF}}
\fancyfoot[RO]{\footnotesize{\sffamily{1--\pageref{LastPage} ~\textbar  \hspace{2pt}\thepage}}}
\fancyfoot[LE]{\footnotesize{\sffamily{\thepage~\textbar\hspace{3.45cm} 1--\pageref{LastPage}}}}
\fancyhead{}
\renewcommand{\headrulewidth}{0pt} 
\renewcommand{\footrulewidth}{0pt}
\setlength{\arrayrulewidth}{1pt}
\setlength{\columnsep}{6.5mm}
\setlength\bibsep{1pt}

\makeatletter 
\newlength{\figrulesep} 
\setlength{\figrulesep}{0.5\textfloatsep} 

\newcommand{\topfigrule}{\vspace*{-1pt}%
\noindent{\color{cream}\rule[-\figrulesep]{\columnwidth}{1.5pt}} }

\newcommand{\botfigrule}{\vspace*{-2pt}%
\noindent{\color{cream}\rule[\figrulesep]{\columnwidth}{1.5pt}} }

\newcommand{\dblfigrule}{\vspace*{-1pt}%
\noindent{\color{cream}\rule[-\figrulesep]{\textwidth}{1.5pt}} }

\makeatother

\twocolumn[
  \begin{@twocolumnfalse}
{\includegraphics[height=30pt]{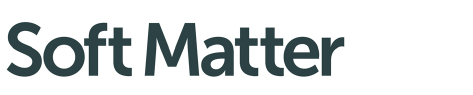}\hfill\raisebox{0pt}[0pt][0pt]{\includegraphics[height=55pt]{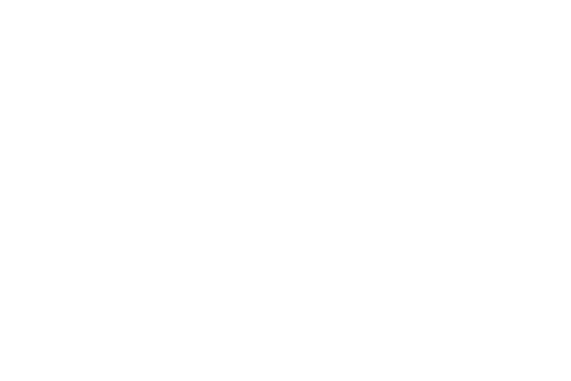}}\\[1ex]
\includegraphics[width=18.5cm]{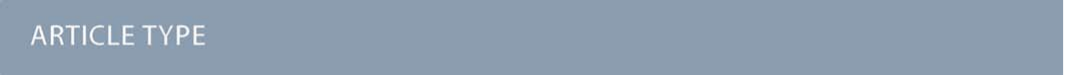}}\par
\vspace{1em}
\sffamily
\begin{tabular}{m{4.5cm} p{13.5cm} }
\includegraphics{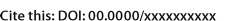} & \noindent\LARGE{\textbf{Hydrodynamics of Multicomponent Vesicles Under Strong Confinement}} \\
\vspace{0.3cm} & \vspace{0.3cm} \\
 & \noindent\large{Ashley Gannon,\textit{$^{a}$} Bryan
  Quaife,\textit{${\ast}${$^{a}$}} and Y.-N.
  Young,\textit{${\ast}$$^{b\ddag}$} } \\
%
\includegraphics{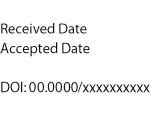} & \noindent\normalsize{
  We numerically investigate the hydrodynamics and membrane dynamics of multicomponent vesicles in two strongly confined geometries. This serves as a simplified model for red blood
  cells undergoing large deformations while traversing 
  narrow constrictions. We propose a new parameterization for the
  bending modulus that remains positive for all lipid phase parameter
  values.	
  For a multicomponent vesicle passing through a stenosis, 
  we establish connections between various properties: lipid phase coarsening, size and flow profile of the lubrication layers, 
  excess pressure, and the tank-treading velocity of
  the membrane. For a multicomponent vesicle passing through a
  contracting channel, we find that the lipid always phase separates so
  that the vesicle is stiffer in the front as it passes through the
  constriction. 
  For both cases of confinement, we find that lipid coarsening is
  arrested under strong confinement and proceeds at a high rate upon
  relief from extreme confinement. The results may be useful for
  efficient sorting lipid domains using microfluidic flows by controlled
  release of vesicles passing through strong confinement.} \\
%
\end{tabular}
 \end{@twocolumnfalse} \vspace{0.6cm}
]

\renewcommand*\rmdefault{bch}\normalfont\upshape
\rmfamily
\section*{}
\vspace{-1cm}


\footnotetext{\textit{$^{a}$~Department of Scientific Computing, Florida State University, Tallahassee, FL, 32306, USA. Email: bquaife@fsu.edu}}
\footnotetext{\textit{$^{b}$~Department of Mathematical Sciences, New Jersey Institute of Technology, Newark, NJ, 07102, USA. Email: yyoung@njit.edu}}




\section{\label{sec:Introduction}Introduction}
Biological membranes, the basic structural units for compartmentalizing
biological systems, 
comprise diverse arrays of
proteins and lipid species. 
These lipids undergo phase separation, forming domains or rafts that
lead to variations in material properties, including bending stiffness.
Synthetic multicomponent vesicles, self-enclosed lipid bilayer membranes
composed of different lipid species, have been used to study the rich
patterns and accompanying morphologies that emerge from elastic
heterogeneity in the membrane. In a quiescent environment, such elastic
heterogeneity gives rise to wrinkling, budding, adhesion, and fusion of
vesicle membranes~\cite{Lowengrub2009_PRE, Li2012_CommMathSci,
Zhao2011_PRE}, closely related to several cellular
processes~\cite{Rauch2000_BiophysJ, Takeda2003_PNAS}.

The hydrodynamics of a single-component vesicle is characterized by
parameters such as the capillary number, reduced volume, and confinement
ratio~\cite{Abreu2014_ACI}. Such characterization is often useful for
inferring fluid and material properties.
For example, a vesicle's reduced volume and viscosity contrast determine
if it undergoes tank-treading, swinging, or tumbling in a linear shear
flow~\cite{nog-gom2005}. Transitions in the vesicle shape occur in
non-linear parabolic flows~\cite{kao-bir-mis2009, dan-vla-mis2009}, and
these shapes include axisymmetric bullets or parachutes, and asymmetric
parachutes. Recently, characterization of vesicles suspended in a
channel or pipe flow~\cite{lyu-che-far-jae-mis-leo2023, aga-bir2020,
qua-gan-you2021, abb-far-nai-ezz-ben-mis2022,
wan-ii-sug-nod-jin-liu-che-gon2023} showed more exotic vesicle
hydrodynamics such as snaking and swirling behaviors in croissant and
slipper shapes.

Under a linear shear flow, a multicomponent vesicle exhibits more exotic
dynamics in both shape and membrane composition than a single-component
vesicle~\cite{soh-tse-li-voi-low2010, Smith2007_JChemPhys,
Cox2015_Nonlinearity, liu-mar-li-vee-low2017, Tusch2018_PRF,
Gera2018_SoftMatter, ger-sal-spa2022}. For example, multicomponent
vesicles under a background flow often exhibit highly complex
morphologies, leading to vesicle budding. The hydrodynamics of these
multicomponent vesicles include phase treading, tumbling with no
viscosity contrast, swinging, budding, and
fission~\cite{soh-tse-li-voi-low2010, wan-du2008, all-ama2006,
ger-sal-spa2022, lip1992, urs-klu-phi2009}. Through numerical
investigations,~\citet{liu-mar-li-vee-low2017} identified several key
dimensionless numbers, including the reduced area, capillary number, and
the floppy-to-stiff ratio to characterize various dynamics of a
two-dimensional multicomponent vesicle in an unbounded linear shear
flow. In addition to the aforementioned numerical studies, these
dynamics have been observed in experiments~\cite{bag-sun2009,
yan-ima-tan2010, yan-ima-tan2008, dre-jah-bob-spa-gop2021}.

Hydrodynamics of vesicles in extreme confinement have been studied as a
model system for red blood cells (RBCs) squeezing through small
capillaries (of sub-micron size in diameter) under a pressure
difference~\cite{Freund2013_PoF, LuPeng2019_PoF, che-lyu-jae-leo2020,
gur-pak-tay-siv-sac2023}. Membrane permeability has been incorporated to
examine the single-component vesicle hydrodynamics. In the absence of an
osmotic gradient, the semipermeable vesicle is affected by water
influx/efflux over a sufficiently long time or under a strong
confinement~\cite{qua-gan-you2021}. Numerical simulations illustrate
that a vesicle with moderate membrane permeability can go through a
strong confinement much more easily, and can restore its water content
within a very short time after its passage through the strong
confinement. These results imply that the membrane permeability may be
inferred from vesicle hydrodynamics under a strong confinement.
Motivated by these results, in this work we seek to investigate how
strong confinement may affect the hydrodynamics of a multicomponent
vesicle. This question is highly relevant to hydrodynamics of cells in a
confinement, and to our knowledge has not been well studied.
\citet{ram-kom-sek-ima2010} reported that confinement reduces the
effective diffusion coefficient of the concentration fluctuation in
multicomponent membranes. How is such a reduced concentration diffusion
coefficient reflected in the hydrodynamics of multicomponent vesicles
under strong confinement?

In this paper, we use numerical simulations and lubrication
analysis~\cite{YoungStone2017_PRF,
mis-wis-ber-key-li-tun-law-per-erd-zha-zha-sun-kal-lam-kon2019} to
investigate the effect of strong confinement on a two-dimensional
multicomponent vesicle. In particular we show how the vesicle's reduced
area and its lipid composition affect the dynamics in two strongly
confined geometries.
The remainder of the paper is organized as follows.
Section~\ref{sec:Formulation} describes the model for the
two-dimensional multicomponent vesicle. We introduce a new
parameterization of the bending modulus that is necessary to avoid
unphysical negative bending stiffness.
Section~\ref{sec:NumericalMethods} describes numerical methods and also
defines the techniques we use to define the excess pressure, lubrication
layer width, and tank-treading velocity. Section~\ref{sec:results}
demonstrates the effects of strong confinement on multicomponent
vesicles. Finally, concluding remarks are made in
Section~\ref{sec:conclusion}.

\section{\label{sec:Formulation}Formulation}
\begin{figure}[h]
  \centering
  \includegraphics[width=0.9\columnwidth]{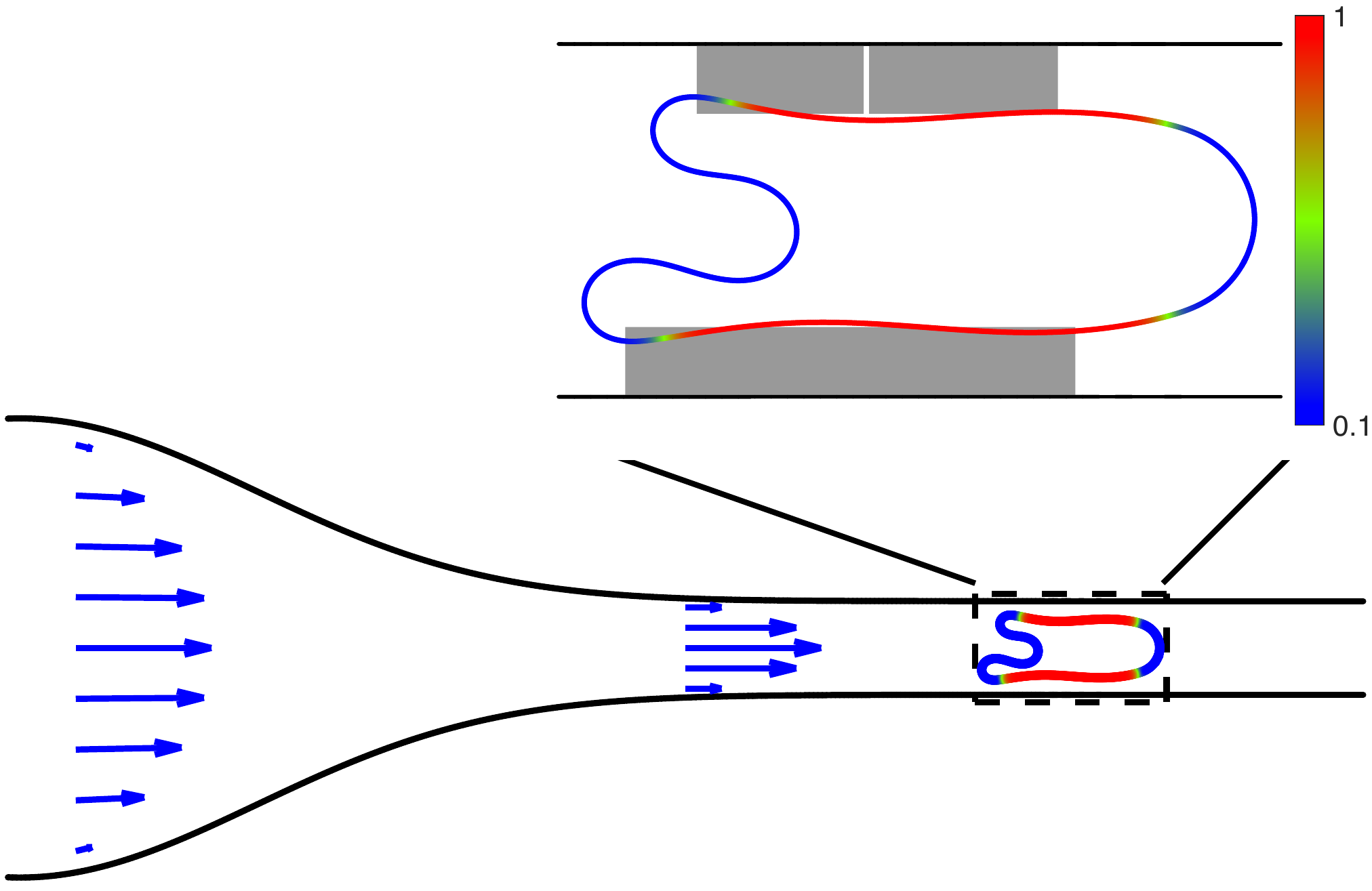}
  \caption{\label{fig:schematic}\small A multicomponent vesicle
  suspended in a strongly-confined stenosis. The color is the
  dimensionless stiffness which varies from $b_{\min} = 0.1$ to
  $b_{\max} = 1$. The vesicle dynamics are determined by a combination
  of phase, bending, and tension energies, and an imposed flow (blue
  arrows). The lubrication layer (gray region), defined in
  Section~\ref{sec:LL}, plays a key role in the vesicle dynamics.}
\end{figure}

We consider a single multicomponent vesicle $\omega$ with boundary
$\gamma$ suspended in a confined geometry $\Omega \subset \RR ^2$ with
boundary $\Gamma$ (Figure~\ref{fig:schematic}). The fluid is assumed to
have zero Reynolds number and is therefore governed by the
incompressible Stokes equations
\begin{align}
  \nabla \cdot \TT = 0 \quad \text{and} \quad \nabla \cdot \uu = 0, 
    \quad \xx \in \Omega \backslash \gamma,
\end{align}
where $\TT = -p\mathbf{I} + \mu\left(\nabla \uu + \nabla \uu^T \right)$
is the hydrodynamic stress tensor, $\uu$ is the velocity, $p$ is the
pressure, and $\mu$ is the fluid viscosity. Across the vesicle membrane,
we require that the velocity is continuous and locally inextensible, and
the membrane and hydrodynamic forces balance
\begin{align}
  \llbracket \uu \rrbracket = 0, \quad 
  \nabla_{\gamma} \cdot \uu = 0, \quad
  \llbracket \TT\nn \rrbracket = \ff, \quad \xx \in \gamma,
\end{align}
where $\ff$ is the total membrane force and $\nn$ is the outward unit
normal of $\gamma$. Parameterizing $\gamma$ as $\xx(s,t)$, where $s$ is
arclength and $t$ is time, the no-slip boundary condition is
$\frac{d\xx}{dt} = \uu(\xx)$. Along the solid wall $\Gamma$, we impose a
Dirichlet boundary condition $\uu(\xx) = \UU(\xx)$, where $\UU$ is a
Hagen–Poiseuille flow at the inlet and outlet
\begin{align}
  \UU(\xx) = U \left(1 - \left(\frac{y}{W}\right)^2 \right), 
    \quad \xx = (x,y) \in \Gamma,
\end{align}
and $y \in [-W,W]$. $\UU$ is zero along the top and bottom of the
channel.

We nondimensionalize the governing equations with a characteristic
length scale $R_0 = 10^{-6}$~m, a maximum bending stiffness $b_{\max} =
10^{-19}$~J, and fluid viscosity $\mu = 5 \times 10^{-2}$~kg/ms. The
resulting bending relaxation time scale is $\mu R_0^3/b_{\max} = 0.5$~s,
the velocity scale is $b_{\max}/\mu R_0^2=2$~$\mu$m/s, the pressure
scale is $b_{\max}/R_0^3 = 10^{-1}$~Pa, and the tension scale is
$b_{\max}/R_0^2 = 10^{-7}$~N/m. The dimensionless parameters regarding
the vesicle properties are the reduced area $\alpha = 4\pi A/L^2$, where
$A$ and $L$ are the vesicle's area and length, respectively, and the
floppy-to-stiffness ratio $\beta = b_{\min}/b_{\max}$. Since we consider
flows in channels, we also define a maximum imposed velocity, $U$, which
sets the capillary number $Ca = U R_{0}^{2} \mu/b_{\max}$, and $W/R_0$
is the dimensionless width of the channel which varies in both examples
we consider. Finally, the Peclet number of the lipid dynamics is taken
to be $Pe = 1$ which results in a diffusion time scale of $2 \times
10^{-4}$~s. Therefore, phase separation occurs at a rate three orders of
magnitude faster than vesicle relaxation. The effect of the Peclet
number has been examined~\cite{liu-mar-li-vee-low2017,
soh-tse-li-voi-low2010}, and it has no qualitative effect on the vesicle
dynamics. From this point onwards, all equations are dimensionless.

\subsection{Constitutive equations\label{subsec:const_eq}}
Using the model first introduced by~\citet{liu-mar-li-vee-low2017}, the
membrane forces depend on the Helfrich energy, line tension, and phase
energy. The phase and bending energies are coupled through the bending
modulus. In particular, the individual energies are
\begin{align}
  E_b &= \frac{1}{2}\int_{\gamma} b(u) \kappa^2 \, ds, \quad
  E_t = \int_{\gamma} \sigma \, ds, \\
  E_p &= \frac{a}{\epsilon}\int_{\gamma}\left(
  f(u) +\frac{\epsilon^2}{2}|\nabla_\gamma u|^2\right) \, ds,
  \label{eqn:PhaseEnergy}
\end{align}
where $u$ is the dimensionless lipid concentration, $b(u)$ is the
lipid-dependent bending modulus, $\sigma$ is the membrane tension,
$\kappa$ is the membrane curvature, and $f(u) = \frac{1}{4}u^2(1-u)^2$
is a double-well potential with local minimums at $u=0$ and $u=1$. The
parameter $\epsilon \ll 1$ sets the size of the transition region of
$u$, and the parameter $a$ is line tension scaled by the characteristic
bending stiffness. All simulations use the parameter values
$\epsilon=100$ and $a=0.04$. The resulting membrane forces are
\begin{align}
  \ff_b &= -(b(u)\kappa \nn)_{ss} -\frac{3}{2}
    \left(b(u) \kappa^2 \ss\right)_s,  \quad
  \ff_t = (\sigma \ss)_s, \\
  \ff_p &= \left(\frac{a}{\epsilon}\left(f(u) -
     \frac{\epsilon^2}{2} u_s^2\right) \ss \right)_s,
\end{align}
where $\ss$ is the unit tangent vector of $\gamma$.

\begin{figure}[h]
  \centering
  \subfigimg[width=0.9\linewidth, clip ]{(a)}{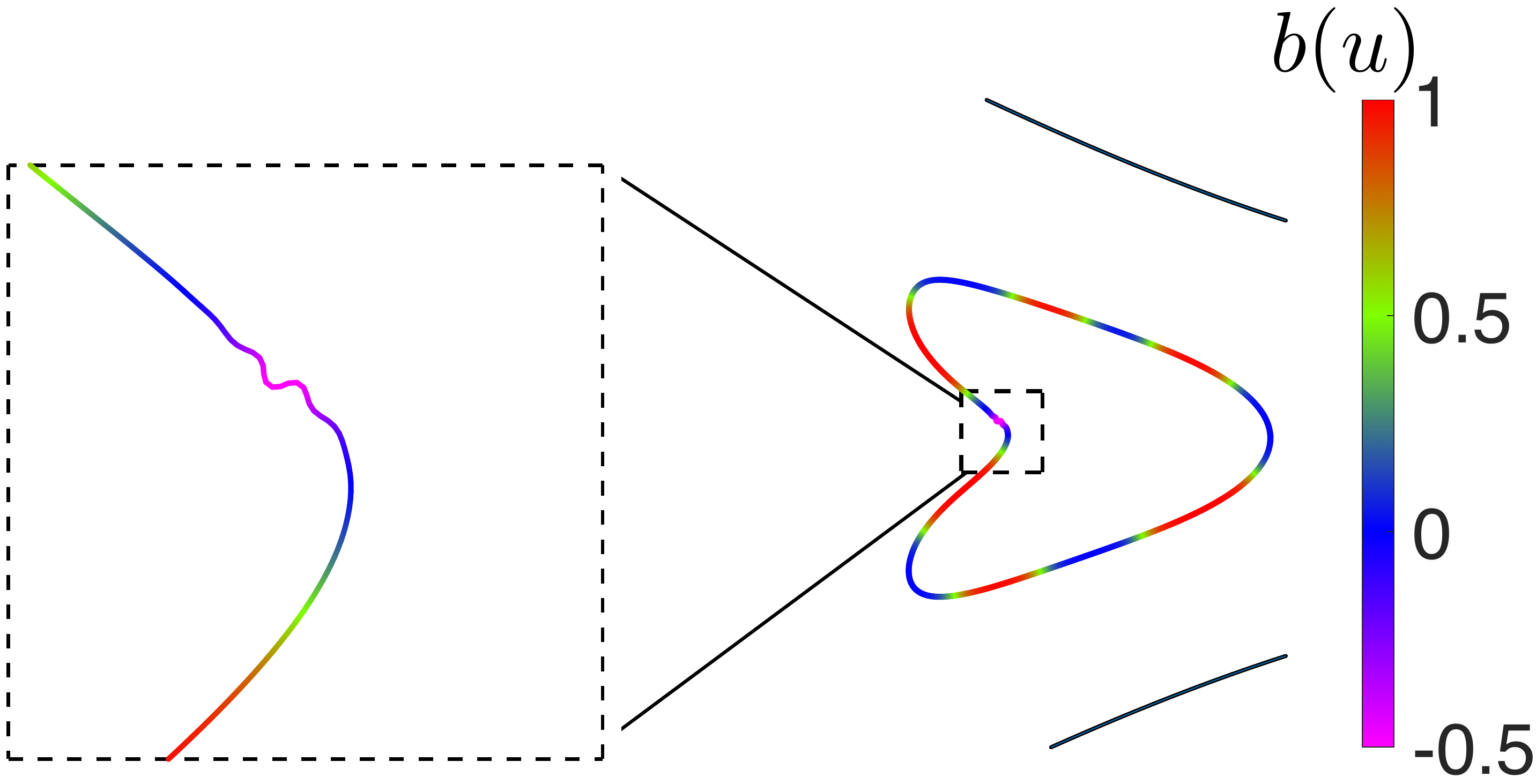} \\
  \subfigimg[width=0.9\linewidth, clip ]{(b)}{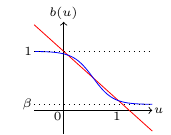}
  \caption{\label{fig:concModels} \small (a) A multicomponent vesicle
  entering a closely-fitting channel (black lines) using the linear
  model in Equation~\eqref{eqn:linearBending}.  The color denotes the
  bending modulus $b(u)$.  An instability is introduced where the
  vesicle bending modulus becomes negative. (b) The linear bending model
  in red (Equation~\eqref{eqn:linearBending}) and the new sigmoid
  bending model in blue (Equation~\eqref{eqn:tanhBending}). Note that
  the new bending model satisfies $b \in (\beta,1)$ even if $u \notin
  [0,1]$.}
\end{figure}

The lipid species $u$ is governed by a fourth-order Cahn-Hilliard
equation that results in the lipids phase separating while conserving
their total mass. To model the variable bending,
\citet{soh-tse-li-voi-low2010} parameterized the bending modulus as
\begin{align}
  \label{eqn:linearBending}
  b(u) = (1-u) + \beta u.
\end{align}
However, since the double-well potential does not have hard walls, the
lipid concentration is not guaranteed to be confined to the interval
$[0,1]$. This is problematic since $b(u) < 0$ if $u > (1 - \beta)^{-1}$,
and such values of $u$ are possible when $\beta \ll 1$, $Ca \gg 1$, or
the vesicle is confined to a narrow region. This behavior is
demonstrated in Figure~\ref{fig:concModels}(a) for a vesicle with a
floppy-to-stiff ratio $\beta = 10^{-1}$ entering a stenosis. Using the
linear bending model in Equation~\eqref{eqn:linearBending}, the lipid
concentration achieves a maximum value of $u \approx 1.5$ which results
in an instability that is formed by the resulting unphysical negative
bending stiffness. To rectify this issue, we parameterize the bending
modulus with the sigmoid function
\begin{align}
  b(u) = \frac{\beta-1}{2} \tanh\left(3\left(u-\frac{1}{2} 
    \right)\right) + \frac{\beta + 1}{2}.
  \label{eqn:tanhBending}
\end{align}
This parameterization maps the local minimums of the double-well
potential close to $\beta$ and $1$, but more importantly, $b(u)$ remains
bounded in $(\beta,1)$ even when $u \notin [0,1]$
(Figure~\ref{fig:concModels}(b)). We compare this new bending model against the linear model for 
a phase-treading and tank treading vesicle in Section~\ref{sec:validation}.


\subsection{\label{sec:NumericalMethods}Numerical Methods}
We use a high-order integral equation formulation to resolve the complex
vesicle shapes and long-range hydrodynamic interactions. The velocity
field $\uu$ at $\xx \in \Omega$ is written as a combination of a
single-layer potential and a double-layer potential
\begin{align}
  \label{eqn:LPrep}
  \uu(\xx) = S[\ff](\xx) + D[\eeta](\xx),
\end{align}
where
\begin{align}
  S[\ff](\xx) &= \frac{1}{4\pi\mu} \int_{\gamma} \left(
    -\log \rho \mathds{I} + \frac{\rr \otimes \rr}{\rho^2} \right)
    \ff(\yy) \, ds_\yy, \\
    D[\eeta](\xx) &= \frac{1}{\pi} \int_{\Gamma} 
      \frac{\rr \cdot \nn}{\rho^2} 
      \frac{\rr \otimes \rr}{\rho^2} \eeta(\yy) \, ds_\yy,
\end{align}
$\rr = \xx - \yy$, $\rho = |\rr|$, and $\mathds{I}$ is the $2 \times 2$
identity matrix. The corresponding pressure is
\begin{align}
  \label{eqn:pressure}
  p(\xx) = \frac{1}{2\pi} \int_{\gamma} \frac{\rr \cdot \ff}{\rho^2} \, ds_\yy + 
    \frac{1}{\pi} \int_{\Gamma} \frac{1}{\rho^2} 
    \left(\mathds{I} - 2\frac{\rr \otimes \rr}{\rho^2} \right) 
    \nn \cdot \eeta \, ds_\yy.
\end{align}

To avoid tangling, the vesicle membrane is parameterized in terms of the
$\theta$-$L$ variables~\cite{hou-low-she1994}, where $L$ is the fixed
vesicle length and $\theta$ is the angle between the tangent vector and
the positive $x$-axis. This requires decomposing the velocity field on
$\gamma$ into a normal velocity $V$ and tangential velocity $T$. Then,
the no-slip boundary condition for the vesicle velocity is
\begin{align}
  \frac{d\xx}{dt} = V \nn + T \ss + D[\eeta](\xx), \quad \xx \in \gamma.
\end{align}
The boundary condition on $\Gamma$ is imposed by requiring that $\eeta$
satisfies
\begin{align}
  \UU(\xx) = -\frac{1}{2}\eeta(\xx) + 
    S[\ff](\xx) + D[\eeta](\xx), \quad \xx \in \Gamma.
  \label{eqn:DLP_BIE}
\end{align}

The vesicle and solid walls are discretized at a set of collocation
points. The single-layer potential is approximated using quadrature for
weakly-singular integrands, while the double-layer potential is
approximated using the spectrally accurate trapezoid rule. A high-order
quadrature method resolves the hydrodynamic interactions between the
vesicle and the solid wall~\cite{qua-bir2014}. Arclength derivatives are
computed with spectral accuracy using Fourier methods.  The resulting
linear system is solved using GMRES. Finally, time stepping is carried
out using the second-order Adams-Bashforth method.

\subsection{Validation}
\label{sec:validation}
\begin{figure}[h]
  \centering
  \includegraphics[width=0.9\columnwidth]{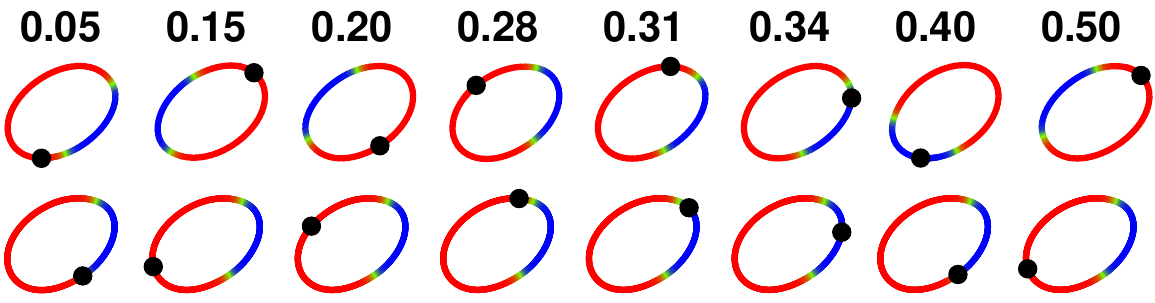}
  \caption{\label{fig:treading} \small A multicomponent vesicle in a
  linear shear flow undergoes phase-treading (top) and tank-treading
  (bottom) with the new sigmoid bending model in
  Equation~\eqref{eqn:tanhBending}. The dimensionless times are in the
  title. The red region is stiff and the blue region is floppy. The
  simulations agree with the results
  in~\citet{liu-mar-li-vee-low2017}.}
\end{figure}

We validate the new sigmoid bending model in
Equation~\eqref{eqn:tanhBending} by reproducing dynamics reported in the
literature. Figure~\ref{fig:treading} shows a phase treading (top) and a
tank-treading (bottom) multicomponent vesicle in an unconfined linear
shear flow. The simulations align with the results reported
by~\citet{liu-mar-li-vee-low2017} (cf.~Figure 3 and Figure 4).

\begin{figure*}[h]
  \centering
  \includegraphics[width=0.9\linewidth]{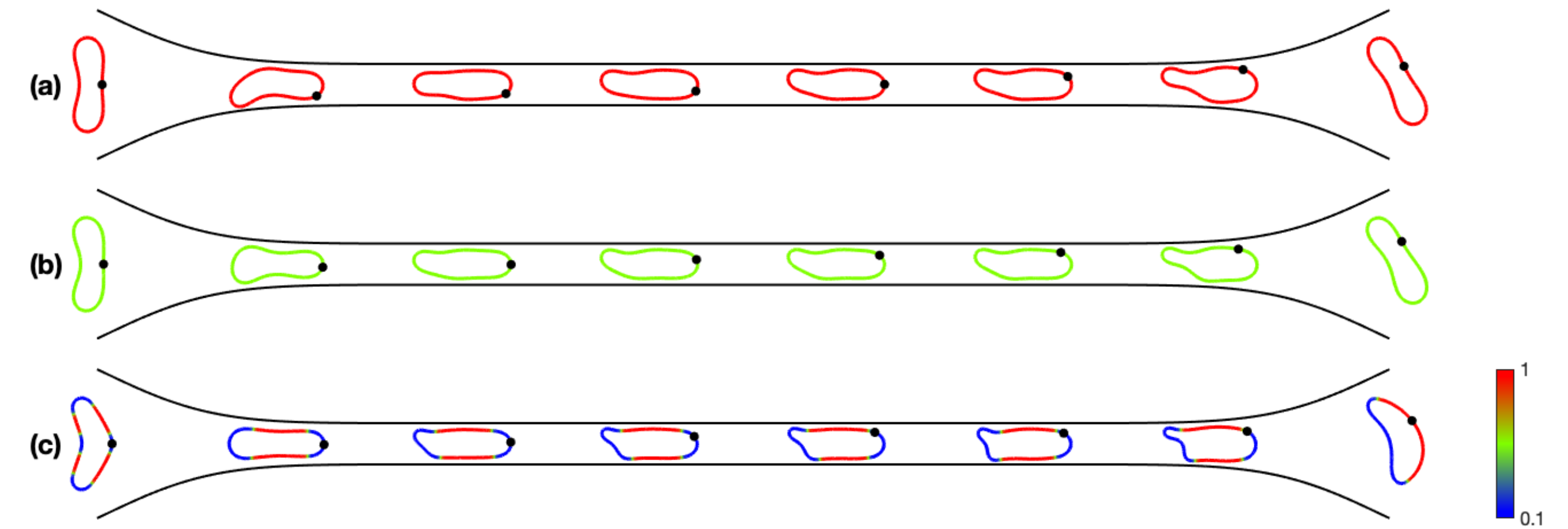}
  \caption{\label{fig:RA6} \small A vesicle passing through the stenosis
  geometry. The vesicle's reduced area is $\alpha = 0.6$ and the
  capillary number is $Ca = 0.25$. The vesicles are: (a) stiff
  single-component; (b) floppy single-component; (c) multicomponent.}
\end{figure*}

\subsection{\label{sec:LL} Lubrication layers and excess pressure}
The lubrication layer between a vesicle and the solid wall plays a
critical role in the dynamics of a vesicle under strong confinement.
Therefore, throughout Section~\ref{sec:results}, we calculate and
analyze the flow inside the lubrication layers between the vesicle and
the channel walls to elucidate the correlation between hydrodynamics and
membrane dynamics. To define the size of the top lubrication layer, we
let $d(\xx,\Gamma_\mathrm{top})$ be the minimum distance between a point
on the vesicle, $\xx \in \gamma$, and the top half of the confining
geometry. The size of the bottom lubrication layer is defined similarly.
We find all local minimums of $d(\xx,\Gamma_\mathrm{top})$, with the
condition that $d(\xx,\Gamma_\mathrm{top}) < \delta$, where $\delta$ is
a parameter. If no such local minimums exist, then the vesicle is too
far from $\Gamma_\mathrm{top}$ to define a lubrication layer. If there
are two or more local minimums, the left-most, $\xx_L$, and right-most,
$\xx_R$, local minimums form the start and end of the lubrication layer.
If there is only one local minimum, then $\xx_L$ and $\xx_R$ are the
points on $\gamma$ to the left and right of this local minimum with the
condition that $d(\xx_L,\Gamma_\mathrm{top}) =
d(\xx_R,\Gamma_\mathrm{top}) = \delta$. Then, the segment
$\gamma_{\mathrm{layer}} \subset \gamma$ consists of all points $\xx \in
\gamma$ between $\xx_L$ and $\xx_R$ with the condition that
$d(\xx,\Gamma_\mathrm{top}) < \delta$. Finally, the lubrication layer
width is
\begin{align}
  w_\mathrm{top} = \frac{1}{|\gamma_{_\mathrm{layer}}|} 
    \int_{\gamma_{_\mathrm{layer}}} d(\xx,\Gamma_\mathrm{top}) \, ds.
\end{align}
The region that defines the top and bottom lubrication layers is
illustrated in Figure~\ref{fig:schematic}, and the height of the gray
regions denote the lubrication layer thicknesses. In this example, the
top layer is broken across two different regions.

A difference between the size of the top and bottom lubrication layers
indicates that the vesicle membrane shape is asymmetric with respect to
the center of the channel, and this results in tank-treading
behavior~\cite{aga-bir2020}. To determine if a vesicle is tank-treading,
we calculate the tangential velocity at several points on $\gamma$, and
if this value is constant, this is the tank-treading velocity.

We also report the excess pressure for the stenosis geometry and relate
it to the size of the lubrication layers. The excess pressure is the
additional pressure that is required to pass the vesicle through the
geometry when compared to the pressure necessary if the vesicle were
absent. Average pressures are calculated using
Equation~\eqref{eqn:pressure} along cross-sections that are parallel to
the flow direction near the inlet and outlet of the stenosis.

\section{\label{sec:results}Hydrodynamics of a multicomponent vesicle
under strong confinement}
We consider both a single-component and a multicomponent vesicle
suspended in two geometries: a closely-fitting geometry (stenosis) and a
geometry that slowly contracts to a narrow neck and then quickly widens
(contracting). 
All the multicomponent examples have floppy-to-stiff ratio $\beta = 10^{-1}$.
For the single-component case, we consider two bending stiffnesses:
$b(u) = 1$ and $b(u) = 0.55$. We refer to these cases as a stiff
single-component vesicle and a floppy single-component vesicle,
respectively. The smaller bending stiffness represents the average
bending stiffness of a multicomponent vesicle, when its stiff and floppy
regions cover the same amount of area. Throughout this section we plot
the position of the vesicle in terms of the $x$-coordinate of its center
of mass. For the stenosis geometry in Section~\ref{subsec:Stenosis}, the
narrow region begins at $x=-15$ and ends at $x=15$. For the contracting
geometry in Section~\ref{subsec:Contraction}, the geometry begins to
narrow at $x=4$ and reaches its narrowest point at $x=15$.

\subsection{\label{subsec:Stenosis}A multicomponent vesicle in the
stenosis geometry}

We start by considering a vesicle with reduced area $\alpha = 0.6$
passing through stenosis with a capillary number $Ca = 0.25$.  With the
current setup and $Ca=0.25$, an elliptical vesicle with reduced area
greater than $\alpha = 0.8$ cannot fit through the closely-fitting
geometry. Figure~\ref{fig:RA6} shows six time steps of a (a) stiff
single-component (b) floppy single-component, and (c) multicomponent
vesicle. The lipid distribution of the multicomponent vesicle is
initially random with a mean value of 0.55. In all plots, the color is
the dimensionless bending modulus of the multicomponent vesicle that
remains bounded in $[0.1,1]$ by using the new sigmoid bending modulus in
Equation~\eqref{eqn:tanhBending}. A single Lagrangian point is included
to visualize any tank-treading behavior. There are slight differences
between the three cases, with the most noticeable being that the
multicomponent vesicle has higher curvature in the floppy region.
However, in general, there is little difference between these cases,
principally because these vesicles have a large reduced area. 

\begin{figure*}[h]
  \centering
  \includegraphics[width=0.9\linewidth]{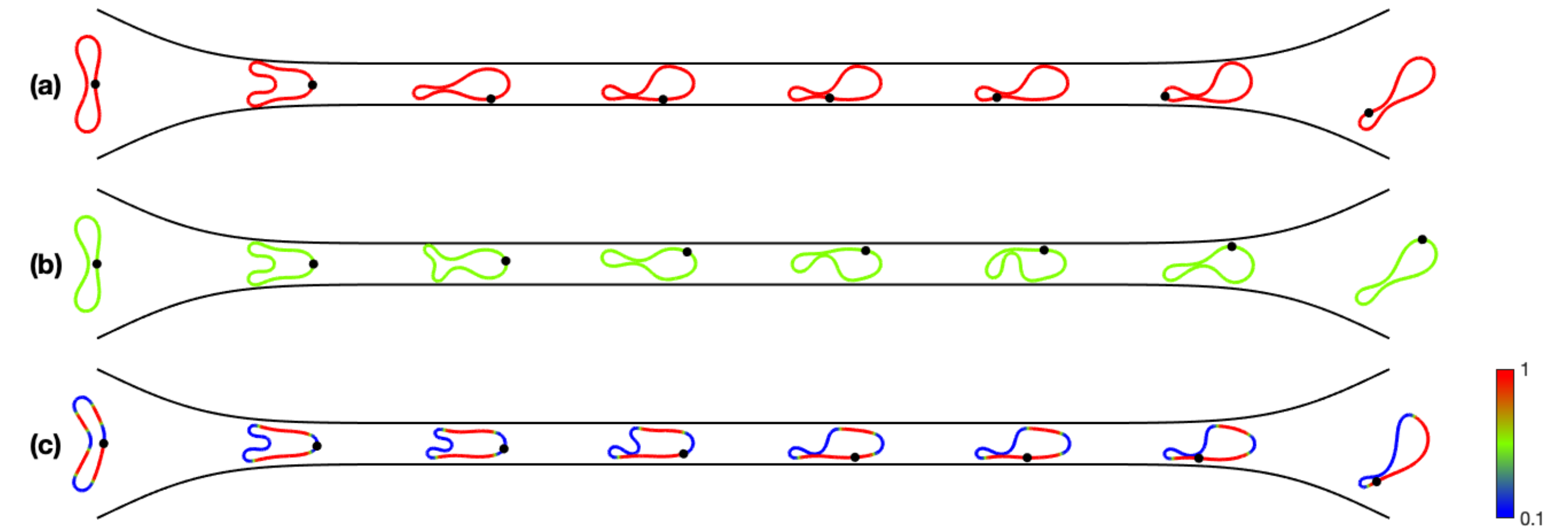}
  \caption{\label{fig:RA4} \small A vesicle passing through the stenosis
  geometry. The vesicle's reduced area is $\alpha = 0.4$ and the
  capillary number is $Ca = 0.25$. The vesicles are: (a) stiff
  single-component; (b) floppy single-component; (c) multicomponent.}
\end{figure*}

\begin{figure*}[ht]
  \centering
  \subfigimg[width=0.3\linewidth]{(a)}{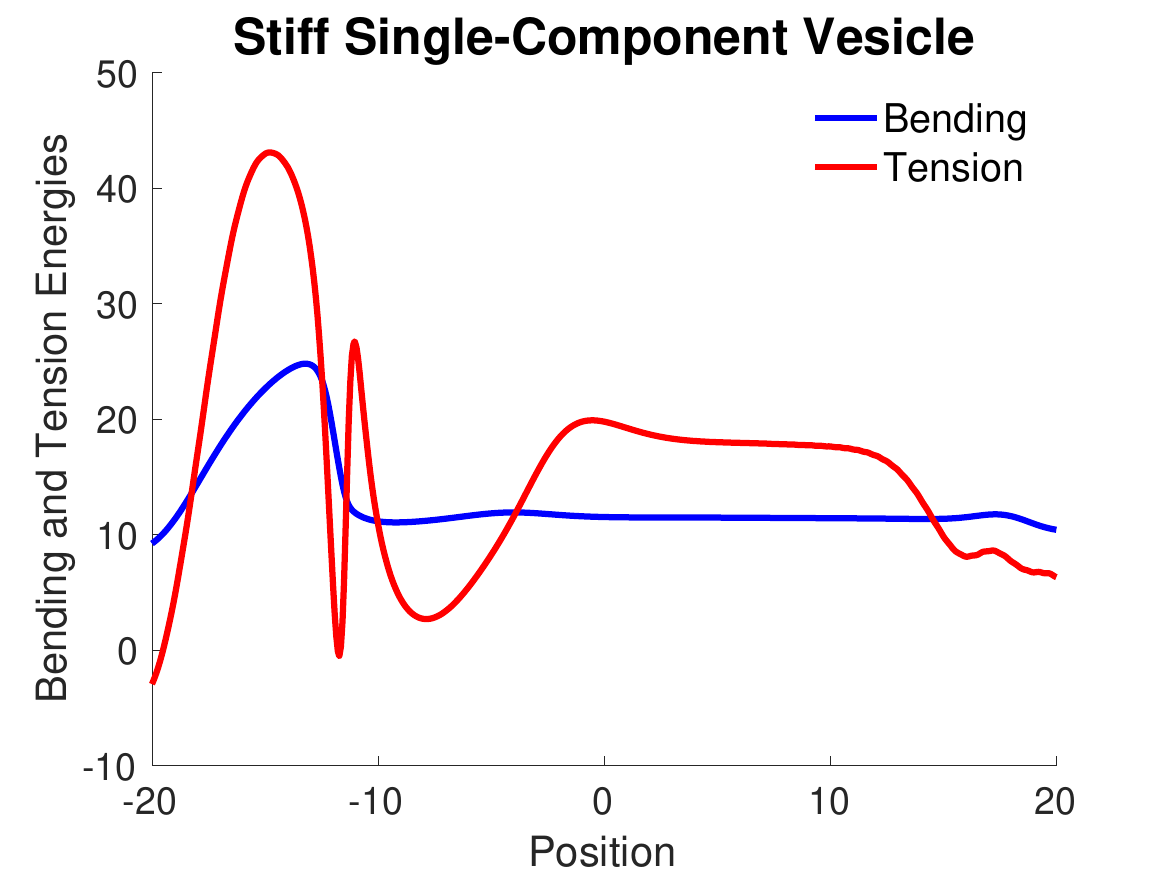}
  \subfigimg[width=0.3\linewidth]{(b)}{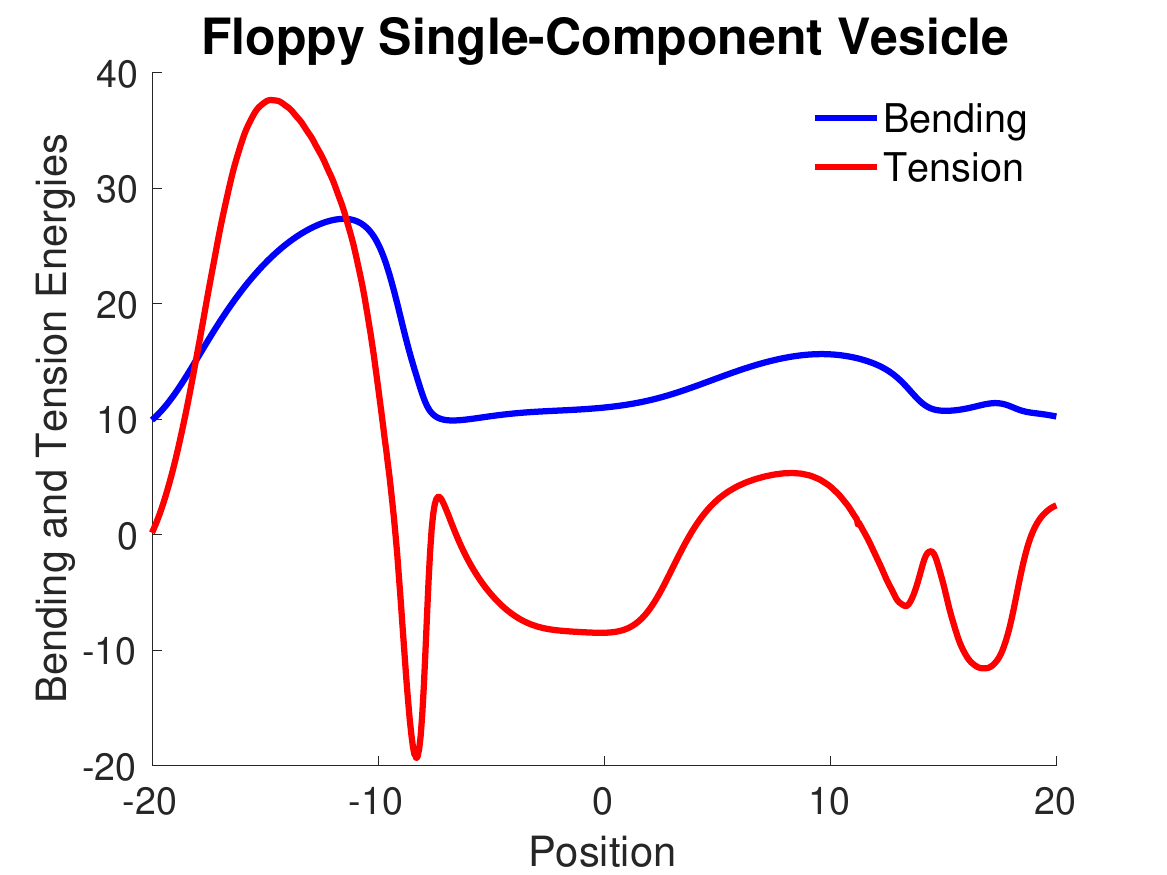}
  \subfigimg[width=0.3\linewidth]{(c)}{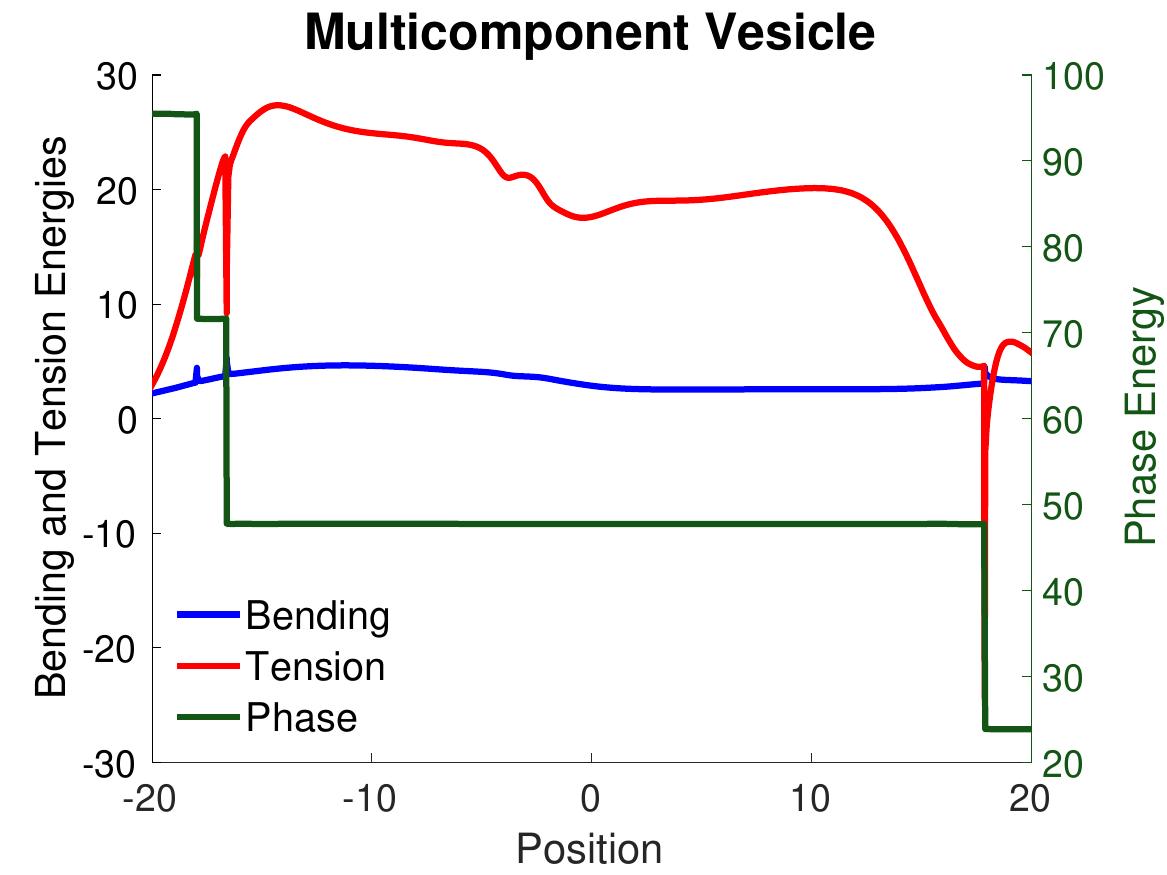}
  \caption{\label{fig:Energy} \small The bending (blue), tension (red),
  and phase (green) energies of a (a) stiff single-component, (b)
  floppy single-component, and (c) multicomponent vesicle. The reduced
  area is $\alpha = 0.4$. Note that the single-component vesicles do not
  have a phase energy.} 
\end{figure*}
\begin{figure*}[ht]
  \centering
  \subfigimg[width=0.3\linewidth]{(a)}{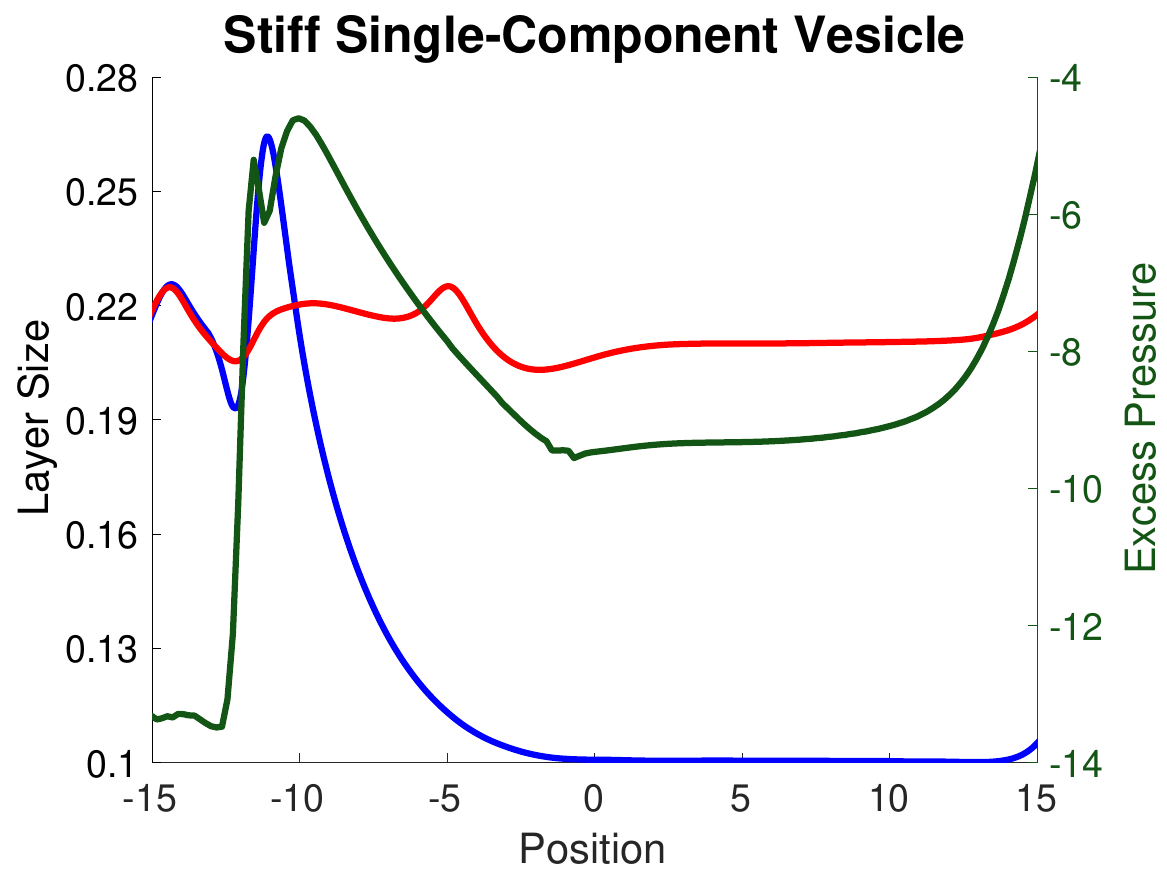}
  \subfigimg[width=0.3\linewidth]{(b)}{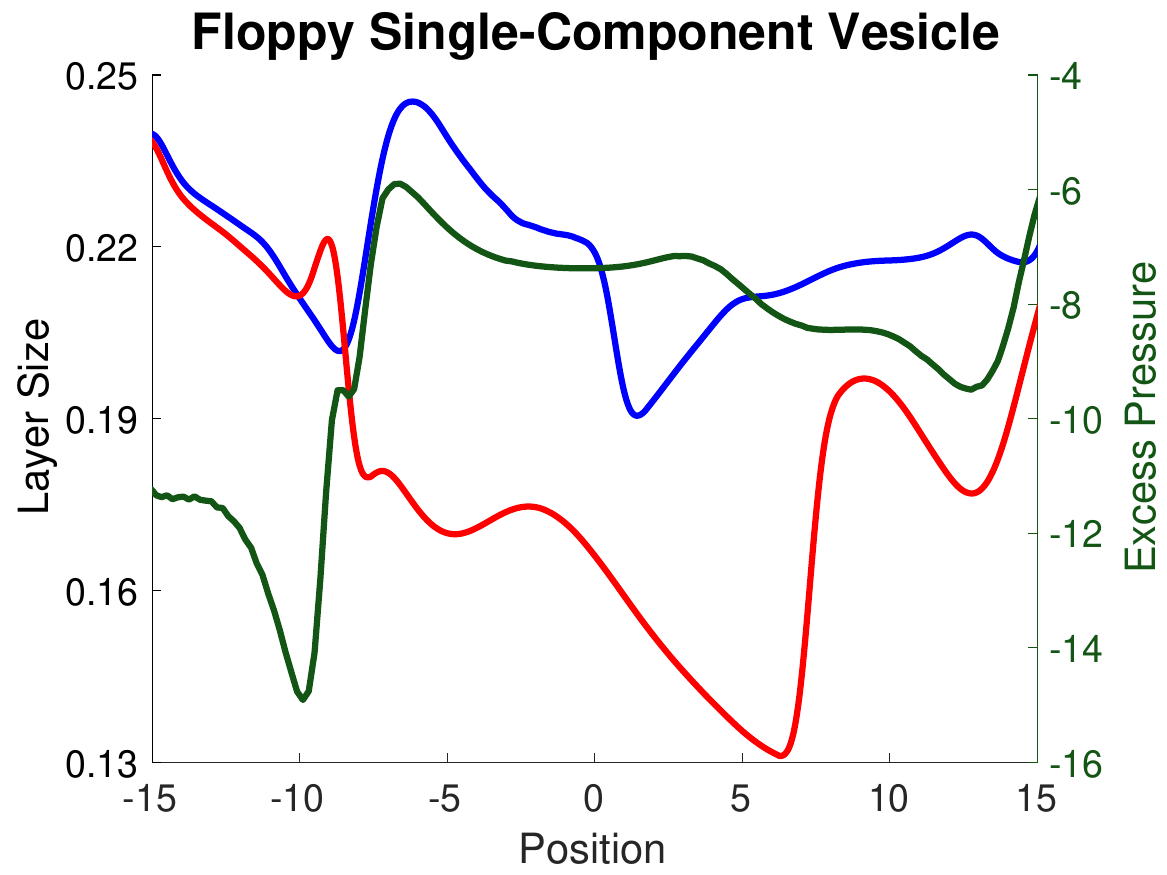}
  \subfigimg[width=0.3\linewidth]{(c)}{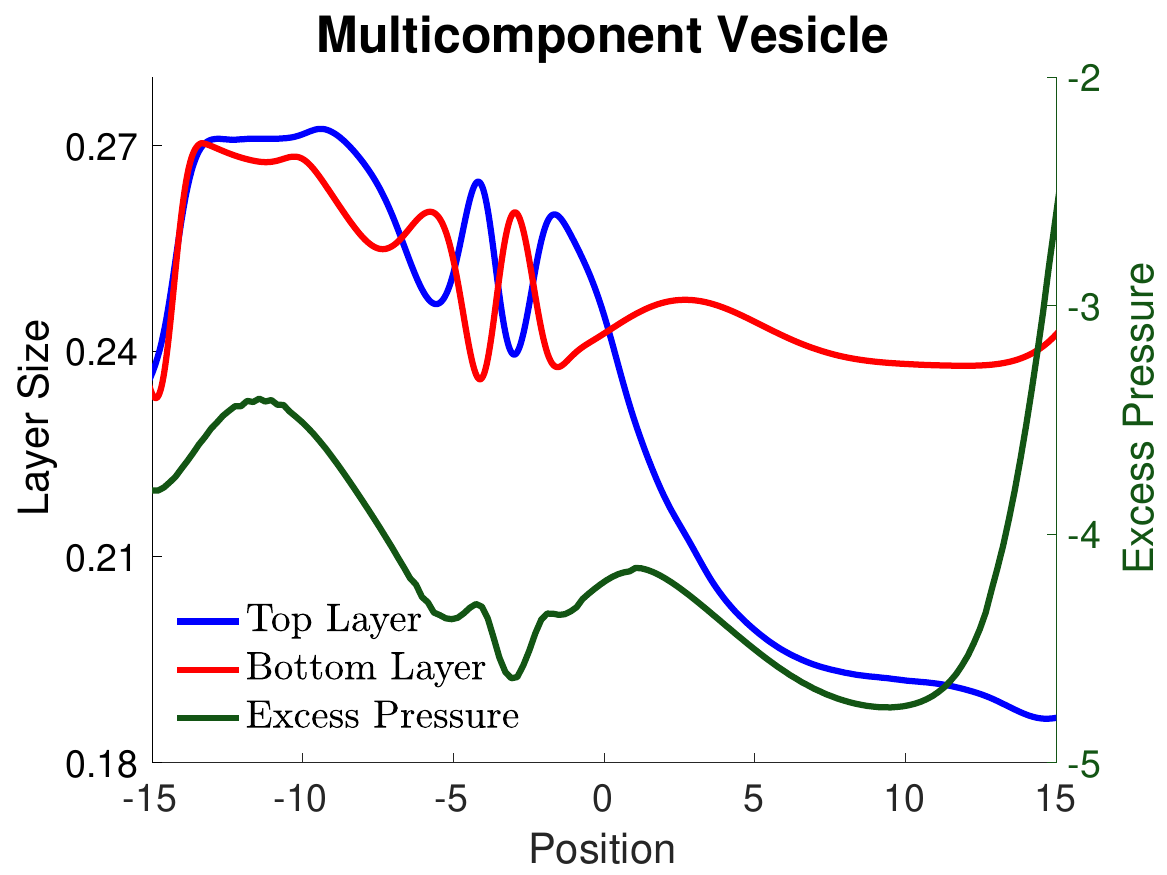}
  \subfigimg[width=0.3\linewidth,trim=0cm 6cm 28cm 0cm,clip=true]{}{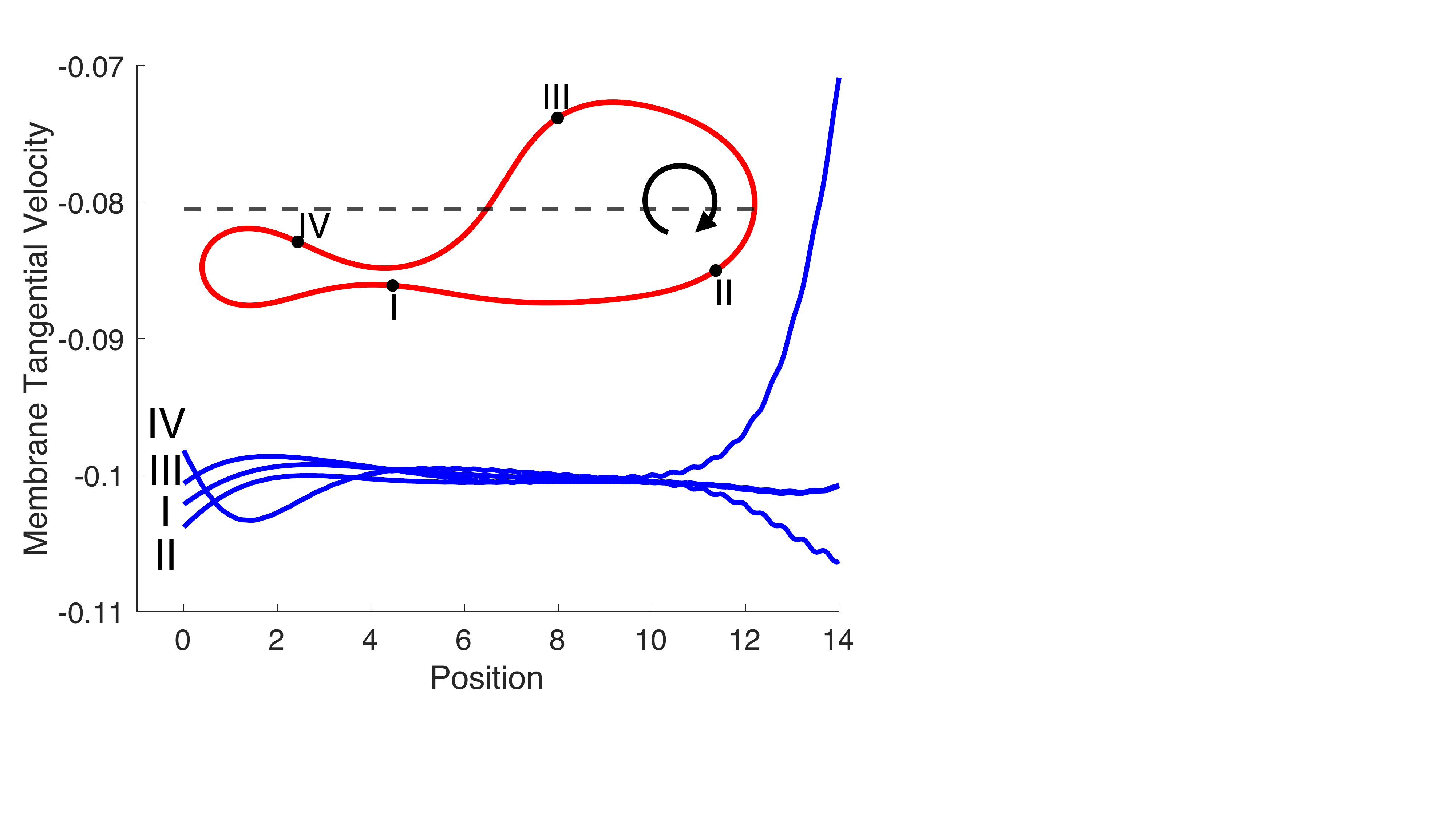}
  \subfigimg[width=0.3\linewidth,trim=0cm 6cm 28cm
  0cm,clip=true]{}{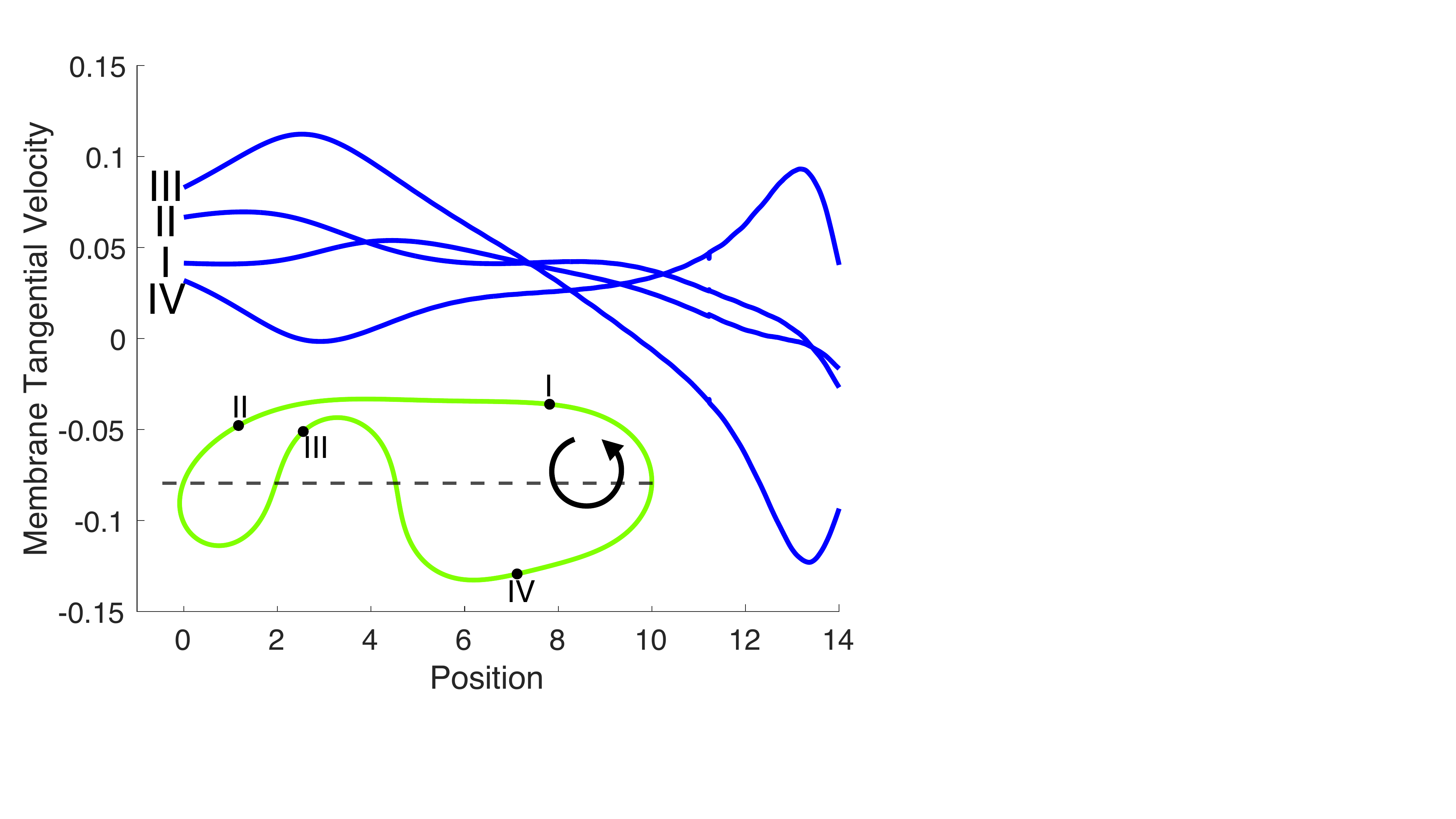}
  \subfigimg[width=0.3\linewidth,trim=0cm 6cm 28cm
  0cm,clip=true]{}{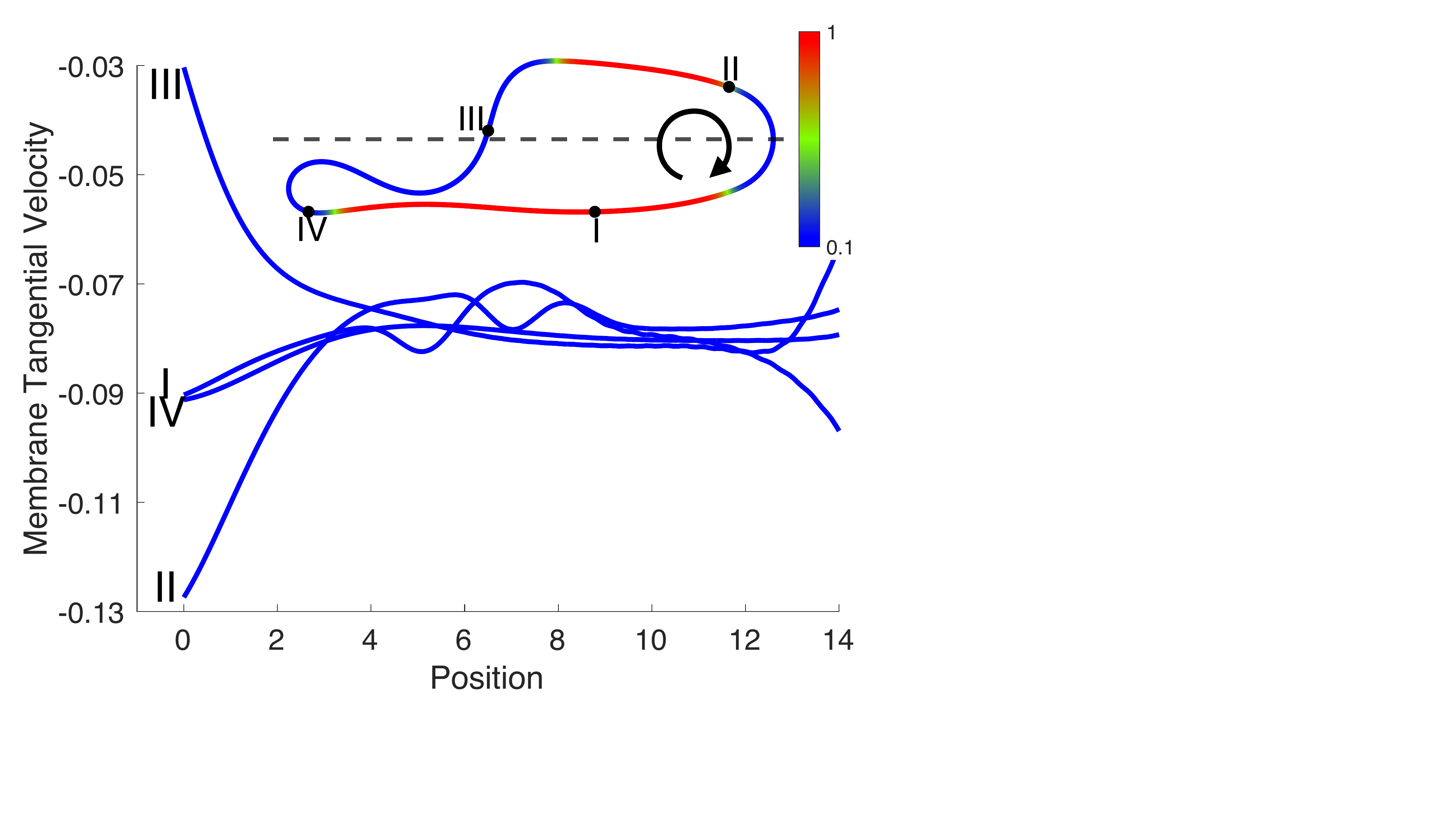}
  \caption{\label{fig:lubricationComposite} \small The lubrication layer
  thicknesses (red and blue), excess pressure (green), and
  tank-treading velocity of four Lagrangian points of a (a) stiff
  single-component vesicle, (b) floppy single-component vesicle, and (c)
  multicomponent vesicle. The reduced area is $\alpha = 0.4$ in all
  three cases. Also illustrated is the vesicle shape and the four
  Lagrangian points at the location indicated by the dash line. The
  black dashed line is the center line of the channel, and the arrow
  denotes the tank-treading direction. The percentage of the vesicle
  membrane that is below the center line is (a) 72\%, (b) 44\%, and (c)
  68\%.}
\end{figure*}

We next consider a vesicle with a smaller reduced area $\alpha=0.4$ with
the same geometry and capillary number. Figure~\ref{fig:RA4} shows the
shape of a (a) stiff single-component, (b) floppy single-component, and
(c) multicomponent vesicle at six locations along the stenosis channel.
Contrary to the vesicles with the larger reduced area
(Figure~\ref{fig:RA6}), the bending modulus has a large effect on the
vesicle shape. We first consider the three different energies---bending,
tension, and phase---for each example. In Figure~\ref{fig:Energy}, we
plot these energies as a function of the vesicle's center. Note that the
single-component vesicles do not have a phase energy. We observe that,
as the vesicle enters the stenosis, it develops regions of high
curvature, and this results in an increase in the bending energy. As
the stiff single-component and multicomponent vesicles reach a
steady-state shape near the middle of the channel, their bending and
tension energies remain nearly constant from $x=0$ until the end of the
stenosis. For the multicomponent vesicle, the phase energy is also
nearly constant in this region. This implies that there is no coarsening
of lipid phase parameter in the middle of a stenosis.

The coexistence of multiple lipid domains in a vesicle under confinement
results in additional interesting physics. We note that in
Figure~\ref{fig:Energy}(c), the lipid species coarsens at three
instances, and this results in a decrease in the phase energy. Each
coarsening corresponds either to two stiff regions overtaking a small
floppy region, or to two floppy regions overtaking a small stiff region.
In the first case, a region that was originally floppy has become stiff,
and this results in a sudden increase in the bending energy. In the
latter case, because the mass of the lipid phase is conserved, the
overtaken stiff region must relocate to the another stiff-floppy
interface, and this also results in a sudden increase in the bending
energy. These increases are visible in Figure~\ref{fig:Energy}(c) as
small upticks in the blue curve. However, in both cases, the vesicle
smooths out these new regions of high curvature, but it takes some time
since the diffusion time scale is several orders of magnitude smaller
than the bending relaxation time scale.

Another difference between the two reduced areas is that the vesicles
with smaller reduced area undergo tank-treading motions. As mentioned
earlier, tank-treading occurs when the top and bottom lubrication layers
differ in size. The size of the lubrication layers for all three cases
of the vesicle with reduced area $\alpha = 0.4$ are in the top half of
Figure~\ref{fig:lubricationComposite}. The bottom of
Figure~\ref{fig:lubricationComposite} shows the tank-treading velocities
of four Lagrangian points. The inset shows the vesicle configuration in
the closely-fitting part of the channel, the black dashed line is the
center line of the channel, the marks are the four Lagrangian points,
and the rounded arrow denotes the direction of rotation. The stiff
single-component and multicomponent vesicles are clearly tank treading.
The direction and speed of the tank-treading depends on how much of
the vesicle membrane is exposed to positive shear (below the center
line) versus how much is exposed to negative shear (above the center
line) \cite{kao-bir-mis2009}. The majority of the membrane of the stiff single-component and
multicomponent vesicles are below the center line, and this results in a
clockwise tank-treading behavior. In contrast, the majority of the
membrane of the floppy single-component vesicle is above the center
line, and the result is counterclockwise tank-treading. In addition, the
tank-treading velocity can be correlated to the percentage of its
membrane that is below the center line. We note, however, that the
floppy single-component vesicle is still undergoing deformations, so the
tangential velocities have not converged to a fixed value.

The sizes of the top and bottom lubrication layers are also related to
the excess pressure. The green curve on the top half of
Figure~\ref{fig:lubricationComposite} shows the excess pressure (with
the axis aligned on the right side of the plot). When the lubrication
layers are large, most notably for the multicomponent vesicle, less
pressure is required to force the vesicle through the geometry, and the
result is a smaller (less negative) excess pressure. In contrast, when
the lubrication layers are small, most notably for the stiff
single-component vesicle, more pressure is required to force the vesicle
through the geometry, and we the result is a larger (more negative)
excess pressure.

\begin{figure}
  \centering
  \subfigimg[width=\columnwidth]{(a)}{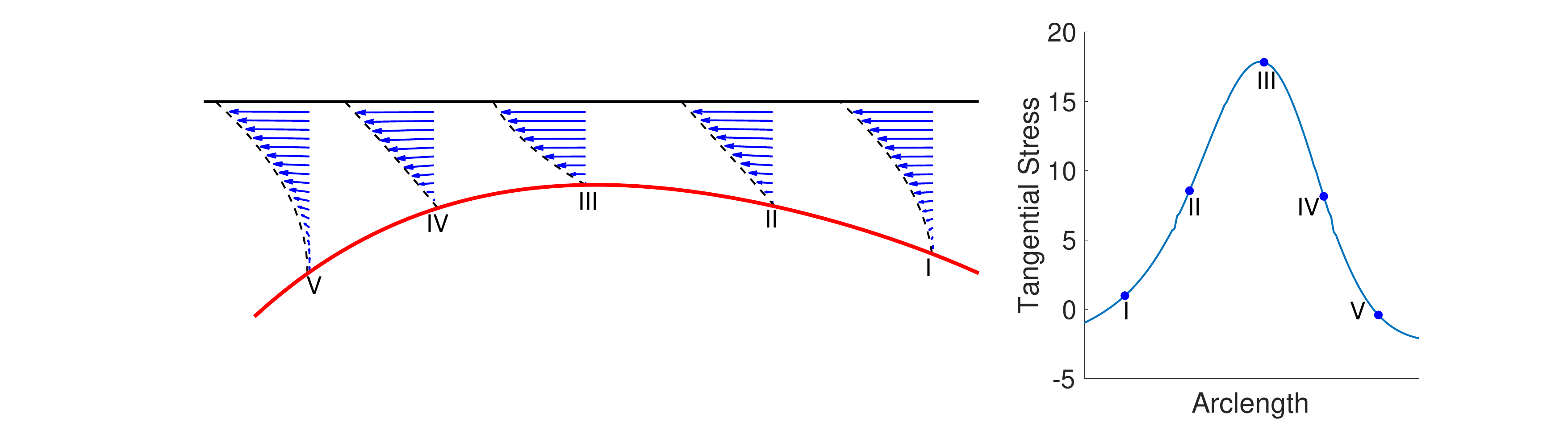}
  \subfigimg[width=\columnwidth]{(b)}{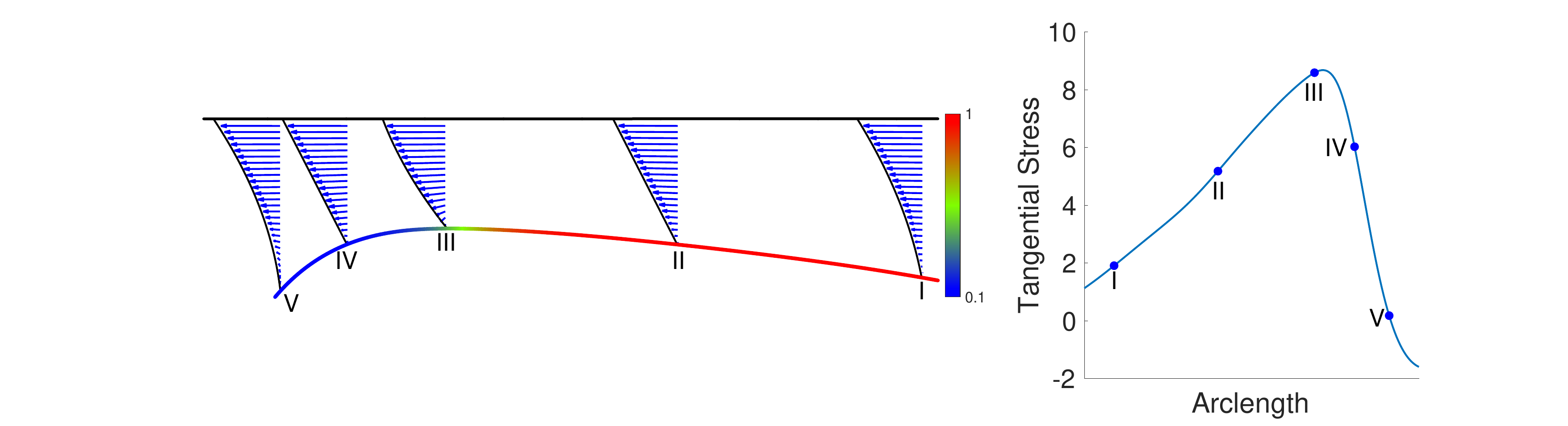}
  \caption{\label{fig:BLvelocities} \small The co-moving velocity field
  in the thin fluid layer between the top of a vesicle and the stenosis
  geometry. The vesicles are (a) stiff single-component; (b)
  multicomponent. Similar velocity profiles between the bottom of the
  vesicle and the stenosis geometry. The profiles are plotted at slices
  where the pressure gradient transitions from a value that is negative,
  zero, positive, zero, and negative. The right plots show the
  tangential stress with the five slices denoted by the marks. Note that
  the reversal in the pressure gradient occurs when the tangential
  stress is sufficiently large.}
\end{figure}

When a vesicle is tank-treading near a solid wall, the flow in the thin
layer of fluid includes both a linear profile and a Poiseuille
profile~\cite{mis-wis-ber-key-li-tun-law-per-erd-zha-zha-sun-kal-lam-kon2019}.
The magnitude of the Poiseuille profile is determined by the pressure
gradient between the vesicle and the solid wall, while the linear
profile is determined by the tangential velocity of the vesicle relative
to the solid wall. Using the definition of the lubrication layer
outlined in Section~\ref{sec:LL}, we consider the flow profile inside
this lubrication layer in Figure~\ref{fig:BLvelocities}. We observe that
the Poiseuille coefficient (quadratic term) can be either positive or
negative, meaning that the pressure gradient in the streamwise direction
in the lubrication layer switches signs. The Poiseuille coefficient is
related to the lateral gradient of the tangential traction,
$\mathbf{F}_\tt$, on the vesicle membrane~\cite{Oron1997_RMP,
Young2014_JFM}
\begin{align}
  \mathbf{F}_{\tt} = -\sigma_s + u_s \frac{\delta E^m}{\delta u}.
\end{align}
Here,
\begin{align}
  \frac{\delta E^m}{\delta u} = \frac{a}{\epsilon} 
    (f'(u) - \epsilon^2 u_{ss}) + \frac{b'(u)}{2} \kappa^2
\end{align}
is the derivative of the membrane energy with respect to the lipid
concentration~\cite{soh-tse-li-voi-low2010}. 

For a single-component vesicle,
the tangential traction is $\mathbf{F}_\tt =-\sigma_s$ (the Marangoni stress), and  the streamwise gradient 
of the Marangoni stress determines the sign of the Poiseuille coefficient in Figure~\ref{fig:BLvelocities}(a).
For a multicomponent vesicle, the Poiseuille coefficient can also change sign around the boundary 
between two lipid domains, where the membrane energy changes sharply between lipid species. 
This correlation between the location of the boundary between lipid domains and the fluid velocity profile in the lubrication layer
implies that, by visualizing the flow
in the lubrication layer and identifying locations where the nonlinear shear flow profile is convex, 
it is possible to identify the location of boundaries between lipid domains as shown in Figure~\ref{fig:BLvelocities}(b).

\subsection{\label{subsec:Contraction} A multicomponent vesicle in the
contracting geometry}
\begin{figure}[h]
  \centering
  \includegraphics[width=\columnwidth]{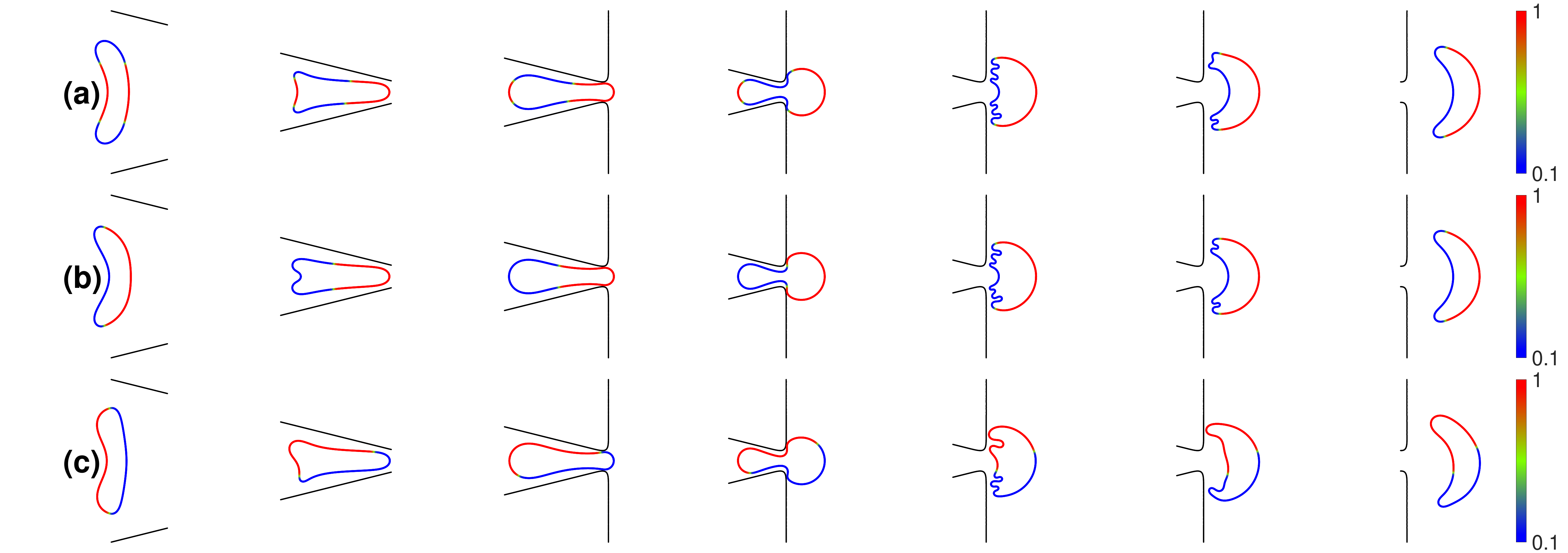}
  \caption{\label{fig:RA6leftRightRand} \small A multicomponent vesicle
  with 50\% floppy region passing through a contracting geometry. The
  vesicle's reduced area is $\alpha = 0.5$ and the capillary number is
  $Ca = 0.25$. The lipid distribution is initialized to be (a) random,
  (b) phase separated with the stiff region leading; (b) random; (c)
  phase separated with the floppy region leading.}
\end{figure}

In this section, we consider a vesicle passing through a channel that
slowly contracts to a  2~$\mu$m wide neck, and then
immediately opens up to a channel that is ten times larger. We
first consider the effect of the initial lipid concentration when the
floppy and stiff regions each make up half of the vesicle membrane. In
Figure~\ref{fig:RA6leftRightRand}, snapshots of vesicles with reduced
area $\alpha = 0.5$ with different initial lipid distributions are
illustrated. The initial distributions are (a) random, (b) stiff in the
front and floppy in the back, and (c) floppy in the front and stiff in
the back. The vesicle with the random initial lipid distribution quickly
phase separates into two floppy and two stiff regions. As seen in the
stenosis geometry, further phase separation is delayed until the vesicle
is no longer under strong confinement. The vesicle with the stiff region
at the front maintains its orientation. In contrast, the vesicle with
the floppy region on the front reorients the floppy region towards the
back of the vesicle. In summary, in all three cases, the vesicle is
orienting itself so that the stiff region passes through the
constriction before the floppy region. Therefore, in subsequent
simulations, we initialize the vesicle's lipid concentration so that the
floppy region is in the back and its stiff region is in the front. The
vesicle shapes develop regions with high curvature in the floppy region
when they first pass through the narrowest part of the geometry, but
once the strong confinement is abruptly removed, these regions of high
curvature are quickly smoothed.

\begin{figure}[h]
  \centering
  \includegraphics[width=\columnwidth]{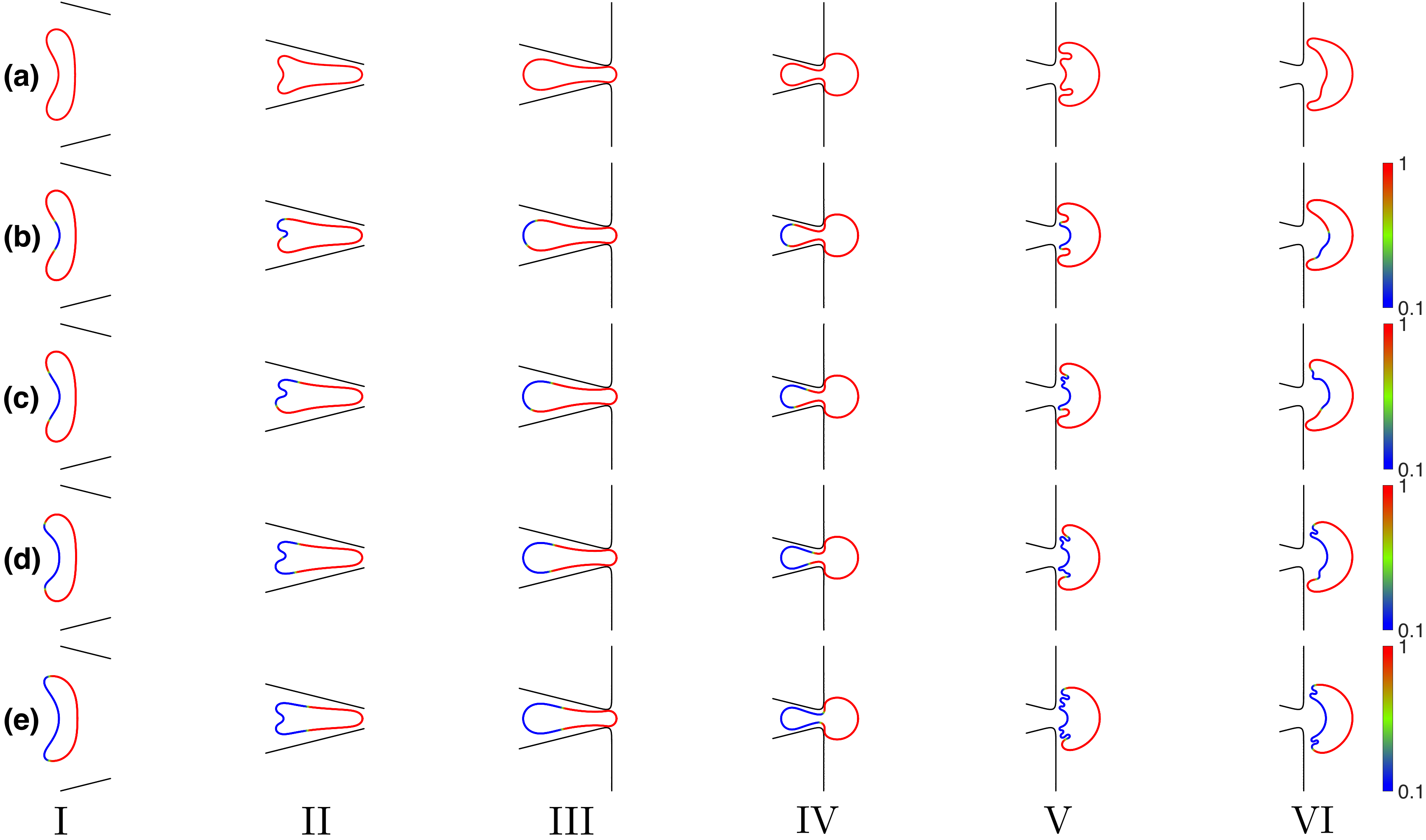}
  \caption{\label{fig:RA5} \small A vesicle passing through a
  contracting geometry. The vesicle's reduced area is $\alpha = 0.5$ and
  the Capillary number is $Ca = 0.25$. (a) A stiff single-component
  vesicle. The percentage of the multicomponent vesicle that is floppy
  is (b) 15\%, (c) 25\%, (d) 35\%, and (e) 45\%. The vesicles with a
  larger percentage of floppy regions pass through the neck earlier than
  the vesicles with less percentage of floppy regions.}
\end{figure}

In addition to the lipid distribution, the percentage of the vesicle
that is floppy plays a role in the vesicle shapes and dynamics.
Therefore, we consider a vesicle with reduced area $\alpha = 0.5$ in the
contracting geometry with various sizes of the floppy region. Motivated
by the last experiment, we initialize the stiff region to be in the
front and the floppy region to be in the back. In each simulation, the
capillary number is $Ca = 0.25$. Figure~\ref{fig:RA5} shows the vesicle
shapes at the same six time steps when the percentage of the floppy
region is (a) 0\% (single-component), (b) 15\%, (c) 25\%, (d) 35\%, and
(e) 45\%. One effect of the size of the floppy region is that the
symmetry of the vesicle shape is broken at smaller percentages. We also
see that floppier vesicles have faster migration speeds, but the effect
is quite small---Table~\ref{tbl:contractingTimes} reports the time the
vesicle first passes through the neck. 

\begin{table}[h]
  \centering
  \begin{tabular}{l|cccccc}
    \hline
    Floppy percentage & 0\% & 15\% & 25\% & 35\% & 45\% & 50\% \\ 
    Dimensionless time & 2.32 & 2.30 & 2.29 & 2.26 & 2.23 & 2.25\\
    \hline
  \end{tabular}
  \caption{\label{tbl:contractingTimes} \small The time required for the
  vesicles to completely pass through the neck.}
\end{table}

\begin{figure}[h]
  \centering
  \includegraphics[width=\columnwidth]{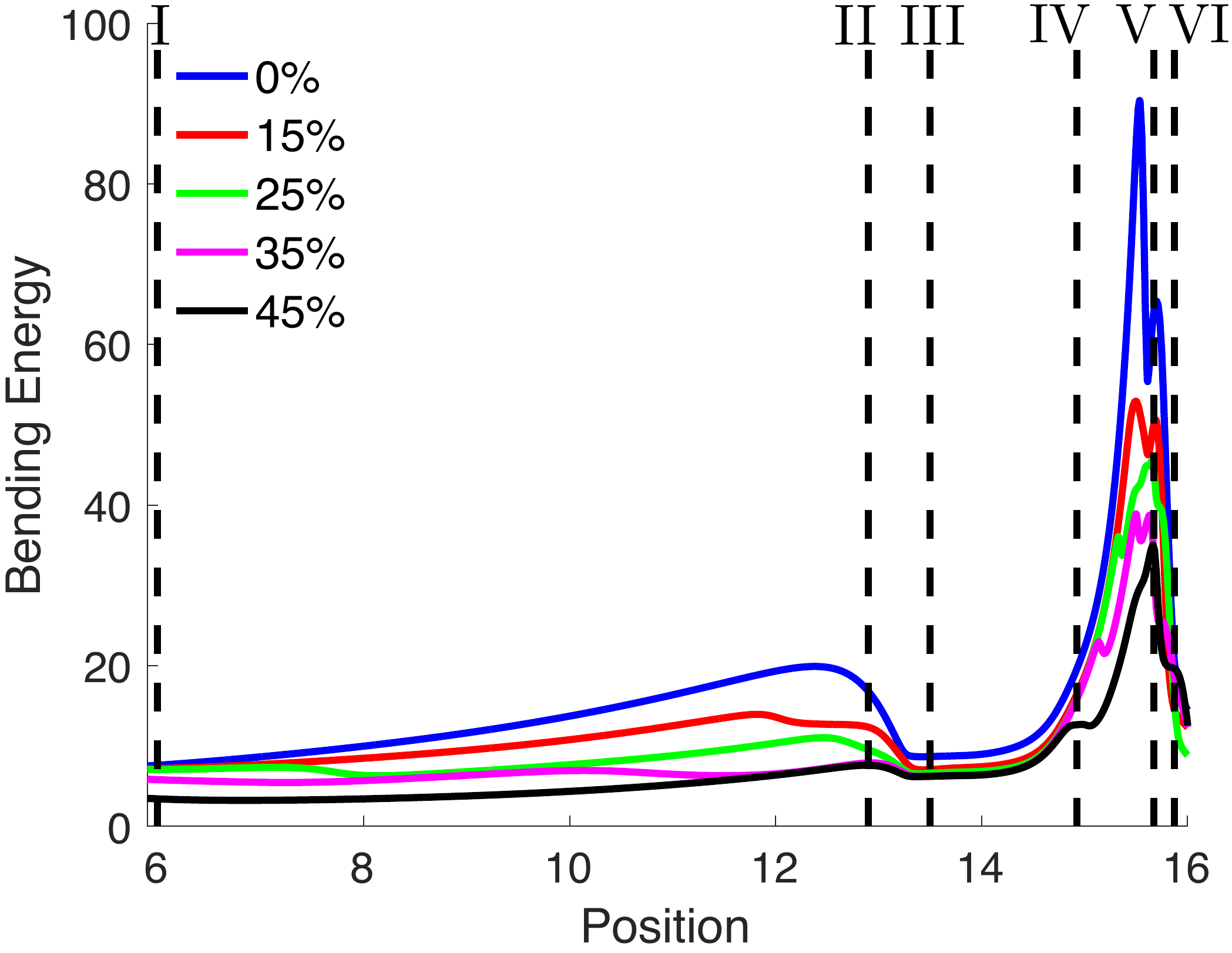}
  \caption{\label{fig:bendingAll} \small The bending energies of the
  single-component and multicomponent vesicles in Figure~\ref{fig:RA5}.
  The vertical dashed lines correspond to the configurations in
  Figure~\ref{fig:RA5}. As the size of the floppy region increases, the
  maximum bending energy decreases.}
\end{figure}

\begin{figure}[h]
  \centering
  \subfigimg[width=\columnwidth]{(a)}{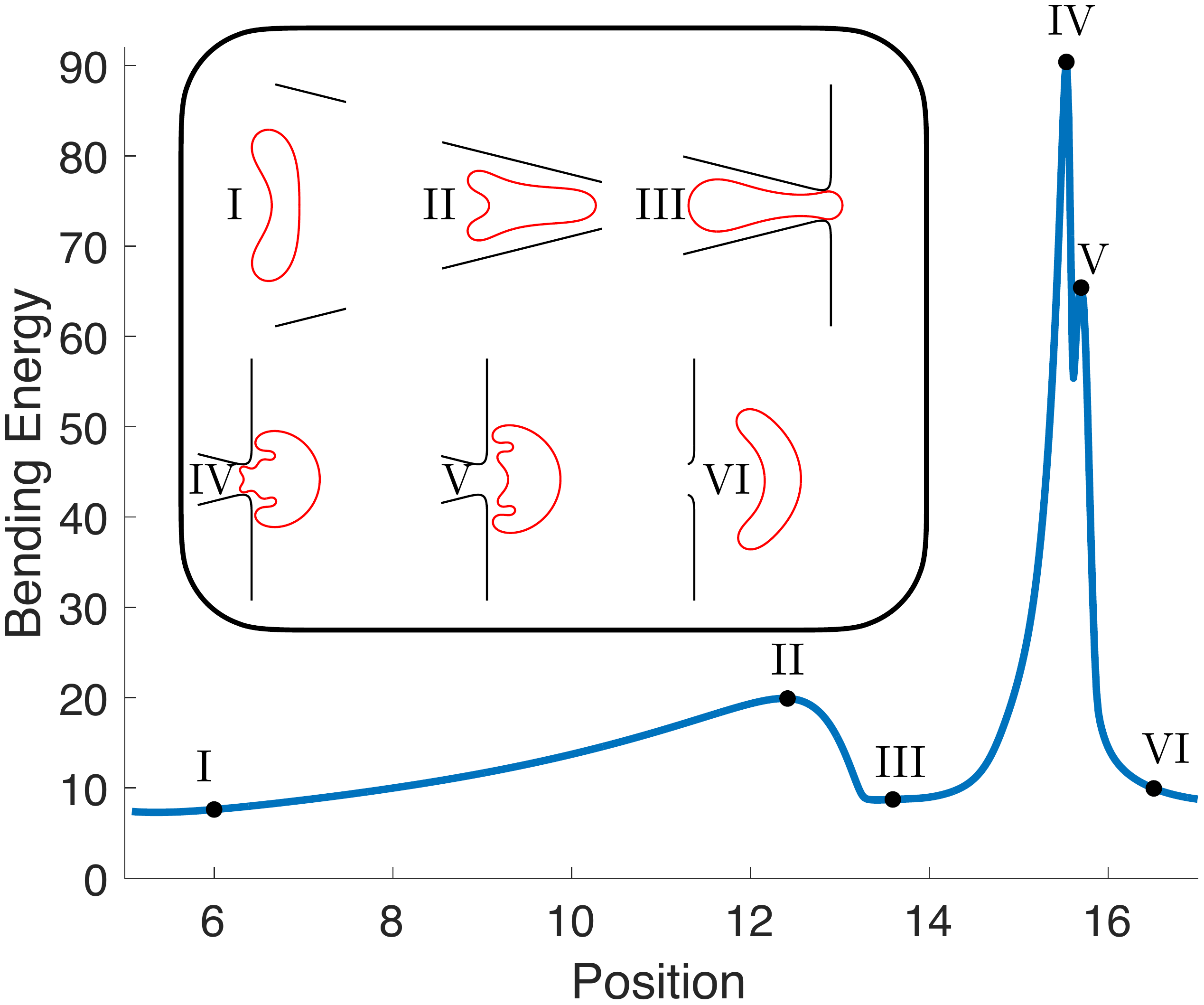}
  \subfigimg[width=\columnwidth]{(b)}{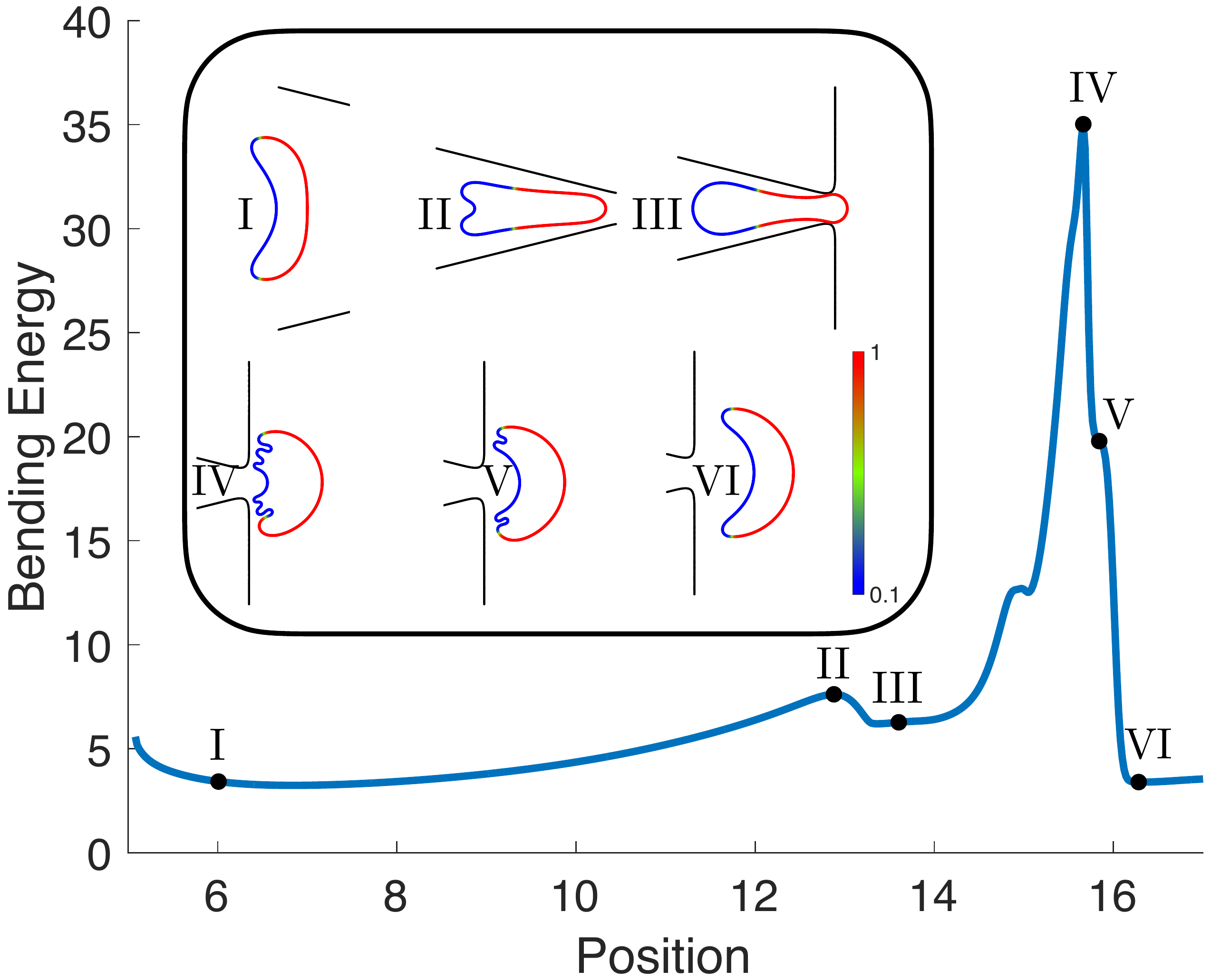}
  \caption{\label{fig:bendingSC_contracting} \small The bending energy
  of (a) a stiff single-component vesicle, and (b) a multicomponent
  vesicle whose boundary is 45\% floppy passing through a contracting
  geometry. The vesicle shape at different locations are included. We
  can clearly see that the sudden increases and decreases in the bending
  energy are due to transitions between lower energy and higher energy
  shapes.}
\end{figure}

We conclude this example by considering the different shapes and
energies that single-component and multicomponent vesicles display.
Since we have initialized the lipid species to initially be phase
separated, the phase energy is nearly constant for all cases, and is
therefore not plotted. The tension energy, on the other hand, increases
as the vesicle approaches the neck, but its general shape is similar for
all five cases, and it too is not plotted. However, the bending energies
show different behaviors for each of the cases
(Figure~\ref{fig:bendingAll}).  The main difference being that the
multicomponent vesicles with a higher percentage of floppy regions have
less bending energy, especially as the vesicle passes through the neck.
We also note that the bending energy undergoes several transient
increases and decreases. These occur when the tail of the vesicle
undergoes transitions from lower energy shapes, such as `C' or `S'
shapes, to high energy shapes, such as `W' shapes.
Figure~\ref{fig:bendingSC_contracting} shows these different vesicle
shapes for the single-component vesicle and multicomponent vesicle that
is 45\% floppy. The shapes at several critical points along the bending
energy are included.

\section{Conclusions \label{sec:conclusion}}
In this work we used numerical simulations to examine the hydrodynamics
of a multicomponent vesicle under strong confinement. The vesicle is an
simplified model for a red blood cell, which has multiple lipid
domains in its membrane and often has to squeeze through a small space.
We refined the phase-field formulation in~\citet{liu-mar-li-vee-low2017}
with a new bending model that uses a sigmoid function for the bending
stiffness to retain positiveness in the bending modulus. The linear
model~\cite{liu-mar-li-vee-low2017} works well for a vesicle in
free-space. However, under strong confinement we find it necessary to
enforce the condition that the dimensionless bending modulus $b(u)$ be
in the range of $[\beta,1]$ ($\beta = b_{\min}/b_{\max}$) to avoid
instability in membrane shapes (see \S~\ref{subsec:const_eq}). We
focused on the hydrodynamics of a multicomponent vesicle in two types of
confinement: a stenosis channel (\S~\ref{subsec:Stenosis}), and a
contracting channel (\S~\ref{subsec:Contraction}). For both confining
geometries we varied the reduced area $\alpha$ and the initial
distribution of lipid domains with a fixed floppy-to-stiff ratio
$\beta=0.1$, a fixed Capillary number $Ca=0.25$, and a fixed Peclet
number of the lipid dynamics $Pe=1$. 

In \S~\ref{subsec:Stenosis} the confinement ratio (defined as $2R_0/W$)
in the stenosis is $1.67$, and in \S~\ref{subsec:Contraction} the
confinement ratio increases to a maximum of $3.60$ at the neck.
Consequently the dependence of the vesicle shape on the capillary number
is expected to be less sensitive according to results
in~\citet{aga-bir2020}. Finally the phase separation is assumed to occur
at a much faster time scale than vesicle relaxation with Peclet number
$Pe=1$.

As a multicomponent vesicle enters and exits a closely-fitted stenosis,
coarsening of the lipid domains is expedited. However, inside the
stenosis, the strong confinement hinders the coarsening as the vesicle
shape and the lipid distribution remain nearly unaltered as the
vesicle moves through the stenosis. Once inside a stenosis, we define
the lubrication layer and its width. We find that tank-treading along
the vesicle inside the stenosis is closely related to the asymmetry in
the lubrication layer width between the top and bottom walls. For
vesicles with a high reduced area the lubrication layer width is
symmetric between the top and bottom walls, and we find no tank-treading
along the membrane. On the other hand, for a vesicle with a lower
reduced area, the lubrication layer width is asymmetric and the vesicle
membrane tanks treads. Furthermore we also find that the excess pressure
(the additional pressure required to pass the vesicle through the
stenosis) is correlated to the size of the lubrication layer: a larger
lubrication layer size results in a smaller excess pressure.

Inside the lubrication layer, the shear flow consists of a quadratic
component (due to pressure gradient in the thin film) and a linear
component (due to membrane motion in the tangential direction). By
analyzing these two components at various locations in the lubrication
layer, we further elucidated the correlation between the location of
boundary between lipid domains and the change of
sign in the pressure gradient in the lubrication layer. These results
may be used to identify the location of boundaries between the lipid
domains via the visualization of the flow in the lubrication layer.

For a multicomponent vesicle squeezing through a constriction, we
simulated the passage of a vesicle with various initial configurations
of lipid domains. We found that, once the vesicle makes its passage, the
front of the vesicle is often stiffer than the back of the vesicle, with
a similar shape of a smaller curvature in the front than the curvature
in the back. This is also observed for a vesicle moving out of
confinement from a stenosis channel. Thus we propose that a strong
confinement followed by a sharp transition to a wider opening can be
used to produce phase-separated multicomponent vesicle with the floppy
lipid species in the back and the stiff lipid species in the front of
the vesicle.

Such asymmetry in lipid domain induced by strong confinement and sudden
release gives rise to a nonuniform tension distribution in the membrane,
with a smaller tension (in magnitude) in the front and a larger tension
(in magnitude) in the back. This resembles the behavior of a red blood cell as it traverses a narrow slit~\cite{LuPeng2019_PoF}. During this process, the membrane tension is primarily negative, resulting in compression that is more pronounced at the rear of the cell than at the front.
\section*{Author Contributions}
Conceptualization: AG, BQ, YNY; methodology: AG, BQ, YNY; validation:
AG, BQ, YNY; formal analysis: AG, BQ, YNY; investigation: AG, BQ, YNY;
data curation: AG, BQ, YNY; writing - original
draft: AG, BQ, YNY ; writing - review and editing: BQ, YNY;
visualization: AG, BQ, YNY; supervision: BQ, YNY; project administration: BQ,
YNY; funding acquisition: BQ, YNY

\section*{Conflicts of interest}
There are no conflicts to declare.

\section*{Acknowledgements}
We thank David Salac  for discussions, and Shuwang Li for discussions and sharing his code for a multicomponent vesicle in free space. 
B.Q.~acknowledges support from the Simons Foundation, Mathematics and Physical
Sciences-Collaboration Grants for Mathematicians, Award No.~527139.
Y.-N.Y.~acknowledges support from NSF (Grants No.~DMS 1614863 and
No.~DMS 195160) and Flatiron Institute, part of Simons Foundation.



\balance



\begin{mcitethebibliography}{40}
\providecommand*{\natexlab}[1]{#1}
\providecommand*{\mciteSetBstSublistMode}[1]{}
\providecommand*{\mciteSetBstMaxWidthForm}[2]{}
\providecommand*{\mciteBstWouldAddEndPuncttrue}
  {\def\EndOfBibitem{\unskip.}}
\providecommand*{\mciteBstWouldAddEndPunctfalse}
  {\let\EndOfBibitem\relax}
\providecommand*{\mciteSetBstMidEndSepPunct}[3]{}
\providecommand*{\mciteSetBstSublistLabelBeginEnd}[3]{}
\providecommand*{\EndOfBibitem}{}
\mciteSetBstSublistMode{f}
\mciteSetBstMaxWidthForm{subitem}
{(\emph{\alph{mcitesubitemcount}})}
\mciteSetBstSublistLabelBeginEnd{\mcitemaxwidthsubitemform\space}
{\relax}{\relax}

\bibitem[Lowengrub \emph{et~al.}(2009)Lowengrub, Ratz, and
  Voigt]{Lowengrub2009_PRE}
J.~S. Lowengrub, A.~Ratz and A.~Voigt, \emph{Physical Review E}, 2009,
  \textbf{79}, 031926\relax
\mciteBstWouldAddEndPuncttrue
\mciteSetBstMidEndSepPunct{\mcitedefaultmidpunct}
{\mcitedefaultendpunct}{\mcitedefaultseppunct}\relax
\EndOfBibitem
\bibitem[Li \emph{et~al.}(2012)Li, Lowengrub, and Voigt]{Li2012_CommMathSci}
S.~Li, J.~Lowengrub and A.~Voigt, \emph{Communications in Mathematical
  Sciences}, 2012, \textbf{10}, 645--670\relax
\mciteBstWouldAddEndPuncttrue
\mciteSetBstMidEndSepPunct{\mcitedefaultmidpunct}
{\mcitedefaultendpunct}{\mcitedefaultseppunct}\relax
\EndOfBibitem
\bibitem[Zhao and Du(2011)]{Zhao2011_PRE}
Y.~Zhao and Q.~Du, \emph{Physical Review E}, 2011, \textbf{84}, 011903\relax
\mciteBstWouldAddEndPuncttrue
\mciteSetBstMidEndSepPunct{\mcitedefaultmidpunct}
{\mcitedefaultendpunct}{\mcitedefaultseppunct}\relax
\EndOfBibitem
\bibitem[Rauch and Farge(2000)]{Rauch2000_BiophysJ}
C.~Rauch and E.~Farge, \emph{Biophysics Journal}, 2000, \textbf{78},
  3036--3047\relax
\mciteBstWouldAddEndPuncttrue
\mciteSetBstMidEndSepPunct{\mcitedefaultmidpunct}
{\mcitedefaultendpunct}{\mcitedefaultseppunct}\relax
\EndOfBibitem
\bibitem[Takeda \emph{et~al.}(2003)Takeda, Leser, Russel, and
  Lamb]{Takeda2003_PNAS}
M.~Takeda, G.~P. Leser, C.~J. Russel and R.~A. Lamb, \emph{Proceedings of the
  National Academy of Science}, 2003, \textbf{100}, 14610--14617\relax
\mciteBstWouldAddEndPuncttrue
\mciteSetBstMidEndSepPunct{\mcitedefaultmidpunct}
{\mcitedefaultendpunct}{\mcitedefaultseppunct}\relax
\EndOfBibitem
\bibitem[Abreu \emph{et~al.}(2014)Abreu, Levant, Steinberg, and
  Seifert]{Abreu2014_ACI}
D.~Abreu, M.~Levant, V.~Steinberg and U.~Seifert, \emph{Advances in Colloid and
  Interface Science}, 2014, \textbf{208}, 129--141\relax
\mciteBstWouldAddEndPuncttrue
\mciteSetBstMidEndSepPunct{\mcitedefaultmidpunct}
{\mcitedefaultendpunct}{\mcitedefaultseppunct}\relax
\EndOfBibitem
\bibitem[Noguchi and Gompper(2005)]{nog-gom2005}
H.~Noguchi and G.~Gompper, \emph{Physical Review E}, 2005, \textbf{72},
  011901\relax
\mciteBstWouldAddEndPuncttrue
\mciteSetBstMidEndSepPunct{\mcitedefaultmidpunct}
{\mcitedefaultendpunct}{\mcitedefaultseppunct}\relax
\EndOfBibitem
\bibitem[Kaoui \emph{et~al.}(2009)Kaoui, Biros, and Misbah]{kao-bir-mis2009}
B.~Kaoui, G.~Biros and C.~Misbah, \emph{Physical Review Letters}, 2009,
  \textbf{103}, 188101\relax
\mciteBstWouldAddEndPuncttrue
\mciteSetBstMidEndSepPunct{\mcitedefaultmidpunct}
{\mcitedefaultendpunct}{\mcitedefaultseppunct}\relax
\EndOfBibitem
\bibitem[Danker \emph{et~al.}(2009)Danker, Vlahovska, and
  Misbah]{dan-vla-mis2009}
G.~Danker, P.~M. Vlahovska and C.~Misbah, \emph{Physical Review Letters}, 2009,
  \textbf{102}, 148102\relax
\mciteBstWouldAddEndPuncttrue
\mciteSetBstMidEndSepPunct{\mcitedefaultmidpunct}
{\mcitedefaultendpunct}{\mcitedefaultseppunct}\relax
\EndOfBibitem
\bibitem[Lyu \emph{et~al.}(2023)Lyu, Chen, Farutin, Jaeger, Misbah, and
  Leonetti]{lyu-che-far-jae-mis-leo2023}
J.~Lyu, P.~G. Chen, A.~Farutin, M.~Jaeger, C.~Misbah and M.~Leonetti,
  \emph{Physical Review Fluids}, 2023, \textbf{8}, L021602\relax
\mciteBstWouldAddEndPuncttrue
\mciteSetBstMidEndSepPunct{\mcitedefaultmidpunct}
{\mcitedefaultendpunct}{\mcitedefaultseppunct}\relax
\EndOfBibitem
\bibitem[Agarwal and Biros(2020)]{aga-bir2020}
D.~Agarwal and G.~Biros, \emph{Physical Review Fluids}, 2020, \textbf{5},
  013603\relax
\mciteBstWouldAddEndPuncttrue
\mciteSetBstMidEndSepPunct{\mcitedefaultmidpunct}
{\mcitedefaultendpunct}{\mcitedefaultseppunct}\relax
\EndOfBibitem
\bibitem[Quaife \emph{et~al.}(2021)Quaife, Gannon, and Young]{qua-gan-you2021}
B.~Quaife, A.~Gannon and Y.-N. Young, \emph{Physical Review Fluids}, 2021,
  \textbf{6}, 073601\relax
\mciteBstWouldAddEndPuncttrue
\mciteSetBstMidEndSepPunct{\mcitedefaultmidpunct}
{\mcitedefaultendpunct}{\mcitedefaultseppunct}\relax
\EndOfBibitem
\bibitem[Abbasi \emph{et~al.}(2022)Abbasi, Farutin, Nait-Ouhra, Ez-Zahraouy,
  Benyoussef, and Misbah]{abb-far-nai-ezz-ben-mis2022}
M.~Abbasi, A.~Farutin, A.~Nait-Ouhra, H.~Ez-Zahraouy, A.~Benyoussef and
  C.~Misbah, \emph{Physical Review Fluids}, 2022, \textbf{7}, 093603\relax
\mciteBstWouldAddEndPuncttrue
\mciteSetBstMidEndSepPunct{\mcitedefaultmidpunct}
{\mcitedefaultendpunct}{\mcitedefaultseppunct}\relax
\EndOfBibitem
\bibitem[Wang \emph{et~al.}(2023)Wang, Ii, Sugiyama, Noda, Jing, Liu, Che, and
  Gong]{wan-ii-sug-nod-jin-liu-che-gon2023}
X.~Wang, S.~Ii, K.~Sugiyama, S.~Noda, P.~Jing, D.~Liu, X.~Che and X.~Gong,
  \emph{Physics of Fluids}, 2023, \textbf{35}, 031910\relax
\mciteBstWouldAddEndPuncttrue
\mciteSetBstMidEndSepPunct{\mcitedefaultmidpunct}
{\mcitedefaultendpunct}{\mcitedefaultseppunct}\relax
\EndOfBibitem
\bibitem[Sohn \emph{et~al.}(2010)Sohn, Tseng, Li, Voigt, and
  Lowengrub]{soh-tse-li-voi-low2010}
J.~S. Sohn, Y.-H. Tseng, S.~Li, A.~Voigt and J.~S. Lowengrub, \emph{Journal of
  Computational Physics}, 2010, \textbf{229}, 119--144\relax
\mciteBstWouldAddEndPuncttrue
\mciteSetBstMidEndSepPunct{\mcitedefaultmidpunct}
{\mcitedefaultendpunct}{\mcitedefaultseppunct}\relax
\EndOfBibitem
\bibitem[Smith and Uspal(2007)]{Smith2007_JChemPhys}
K.~A. Smith and W.~E. Uspal, \emph{Journal of Chemical Physics}, 2007,
  \textbf{126}, 02B610\relax
\mciteBstWouldAddEndPuncttrue
\mciteSetBstMidEndSepPunct{\mcitedefaultmidpunct}
{\mcitedefaultendpunct}{\mcitedefaultseppunct}\relax
\EndOfBibitem
\bibitem[Cox and Lowengrub(2015)]{Cox2015_Nonlinearity}
G.~Cox and J.~Lowengrub, \emph{Nonlinearity}, 2015, \textbf{28}, 773--793\relax
\mciteBstWouldAddEndPuncttrue
\mciteSetBstMidEndSepPunct{\mcitedefaultmidpunct}
{\mcitedefaultendpunct}{\mcitedefaultseppunct}\relax
\EndOfBibitem
\bibitem[Liu \emph{et~al.}(2017)Liu, Marple, Li, Veerapaneni, and
  Lowengrub]{liu-mar-li-vee-low2017}
K.~Liu, G.~R. Marple, S.~Li, S.~Veerapaneni and J.~Lowengrub, \emph{Soft
  Matter}, 2017, \textbf{13}, 3521--3531\relax
\mciteBstWouldAddEndPuncttrue
\mciteSetBstMidEndSepPunct{\mcitedefaultmidpunct}
{\mcitedefaultendpunct}{\mcitedefaultseppunct}\relax
\EndOfBibitem
\bibitem[Tusch \emph{et~al.}(2018)Tusch, Loiseau, H., Khelloufi, Helfer, and
  Viallat]{Tusch2018_PRF}
S.~Tusch, E.~Loiseau, A.-H.~A. H., K.~Khelloufi, E.~Helfer and A.~Viallat,
  \emph{Physical Review Fluids}, 2018, \textbf{3}, 123605\relax
\mciteBstWouldAddEndPuncttrue
\mciteSetBstMidEndSepPunct{\mcitedefaultmidpunct}
{\mcitedefaultendpunct}{\mcitedefaultseppunct}\relax
\EndOfBibitem
\bibitem[Gera and Salac(2018)]{Gera2018_SoftMatter}
P.~Gera and D.~Salac, \emph{Soft Matter}, 2018, \textbf{14}, 7690--7705\relax
\mciteBstWouldAddEndPuncttrue
\mciteSetBstMidEndSepPunct{\mcitedefaultmidpunct}
{\mcitedefaultendpunct}{\mcitedefaultseppunct}\relax
\EndOfBibitem
\bibitem[Gera \emph{et~al.}(2022)Gera, Salac, and Spagnolie]{ger-sal-spa2022}
P.~Gera, D.~Salac and S.~E. Spagnolie, \emph{Journal of Fluid Mechanics}, 2022,
  \textbf{935}, \relax
\mciteBstWouldAddEndPuncttrue
\mciteSetBstMidEndSepPunct{\mcitedefaultmidpunct}
{\mcitedefaultendpunct}{\mcitedefaultseppunct}\relax
\EndOfBibitem
\bibitem[Wang and Du(2008)]{wan-du2008}
X.~Wang and X.~Du, \emph{Journal of Mathematical Biology}, 2008, \textbf{56},
  347--371\relax
\mciteBstWouldAddEndPuncttrue
\mciteSetBstMidEndSepPunct{\mcitedefaultmidpunct}
{\mcitedefaultendpunct}{\mcitedefaultseppunct}\relax
\EndOfBibitem
\bibitem[Allain and Amar(2006)]{all-ama2006}
J.-M. Allain and M.~B. Amar, \emph{The European Physical Journal E}, 2006,
  \textbf{20}, 409--420\relax
\mciteBstWouldAddEndPuncttrue
\mciteSetBstMidEndSepPunct{\mcitedefaultmidpunct}
{\mcitedefaultendpunct}{\mcitedefaultseppunct}\relax
\EndOfBibitem
\bibitem[Lipowsky(1992)]{lip1992}
R.~Lipowsky, \emph{Journal de Physique II}, 1992, \textbf{2}, 1825--1840\relax
\mciteBstWouldAddEndPuncttrue
\mciteSetBstMidEndSepPunct{\mcitedefaultmidpunct}
{\mcitedefaultendpunct}{\mcitedefaultseppunct}\relax
\EndOfBibitem
\bibitem[Ursell \emph{et~al.}(2009)Ursell, Klug, and Phillips]{urs-klu-phi2009}
T.~S. Ursell, W.~S. Klug and R.~Phillips, \emph{Proceedings of the National
  Academy of Science}, 2009, \textbf{106}, 13301--13306\relax
\mciteBstWouldAddEndPuncttrue
\mciteSetBstMidEndSepPunct{\mcitedefaultmidpunct}
{\mcitedefaultendpunct}{\mcitedefaultseppunct}\relax
\EndOfBibitem
\bibitem[Bagatolli and Kumar(2009)]{bag-sun2009}
L.~Bagatolli and P.~B.~S. Kumar, \emph{Soft Matter}, 2009, \textbf{5},
  3234--3248\relax
\mciteBstWouldAddEndPuncttrue
\mciteSetBstMidEndSepPunct{\mcitedefaultmidpunct}
{\mcitedefaultendpunct}{\mcitedefaultseppunct}\relax
\EndOfBibitem
\bibitem[Yanagisawa \emph{et~al.}(2010)Yanagisawa, Imai, and
  Taniguchi]{yan-ima-tan2010}
M.~Yanagisawa, M.~Imai and T.~Taniguchi, \emph{Physical Review E}, 2010,
  \textbf{82}, 051928\relax
\mciteBstWouldAddEndPuncttrue
\mciteSetBstMidEndSepPunct{\mcitedefaultmidpunct}
{\mcitedefaultendpunct}{\mcitedefaultseppunct}\relax
\EndOfBibitem
\bibitem[Yanagisawa \emph{et~al.}(2008)Yanagisawa, Imai, and
  Taniguchi]{yan-ima-tan2008}
M.~Yanagisawa, M.~Imai and T.~Taniguchi, \emph{Physical Review Letters}, 2008,
  \textbf{100}, 148102\relax
\mciteBstWouldAddEndPuncttrue
\mciteSetBstMidEndSepPunct{\mcitedefaultmidpunct}
{\mcitedefaultendpunct}{\mcitedefaultseppunct}\relax
\EndOfBibitem
\bibitem[Dreher \emph{et~al.}(2021)Dreher, Jahnke, Bobkova, Spatz, and
  G\"opfrich]{dre-jah-bob-spa-gop2021}
Y.~Dreher, K.~Jahnke, E.~Bobkova, J.~P. Spatz and K.~G\"opfrich,
  \emph{Angewandte Chemie}, 2021, \textbf{60}, 10661--10669\relax
\mciteBstWouldAddEndPuncttrue
\mciteSetBstMidEndSepPunct{\mcitedefaultmidpunct}
{\mcitedefaultendpunct}{\mcitedefaultseppunct}\relax
\EndOfBibitem
\bibitem[Freund(2013)]{Freund2013_PoF}
J.~B. Freund, \emph{Physics of Fluids}, 2013, \textbf{25}, 110807\relax
\mciteBstWouldAddEndPuncttrue
\mciteSetBstMidEndSepPunct{\mcitedefaultmidpunct}
{\mcitedefaultendpunct}{\mcitedefaultseppunct}\relax
\EndOfBibitem
\bibitem[Lu and Peng(2019)]{LuPeng2019_PoF}
H.~Lu and Z.~Peng, \emph{Physics of Fluids}, 2019, \textbf{31}, 031902\relax
\mciteBstWouldAddEndPuncttrue
\mciteSetBstMidEndSepPunct{\mcitedefaultmidpunct}
{\mcitedefaultendpunct}{\mcitedefaultseppunct}\relax
\EndOfBibitem
\bibitem[Chen \emph{et~al.}(2020)Chen, Lyu, Jaeger, and
  Leonetti]{che-lyu-jae-leo2020}
P.~G. Chen, J.~M. Lyu, M.~Jaeger and M.~Leonetti, \emph{Physical Review
  Fluids}, 2020, \textbf{5}, 043602\relax
\mciteBstWouldAddEndPuncttrue
\mciteSetBstMidEndSepPunct{\mcitedefaultmidpunct}
{\mcitedefaultendpunct}{\mcitedefaultseppunct}\relax
\EndOfBibitem
\bibitem[G\"urb\"uz \emph{et~al.}(2023)G\"urb\"uz, Pak, Taylor, Sivaselvan, and
  Sachs]{gur-pak-tay-siv-sac2023}
A.~G\"urb\"uz, O.~S. Pak, M.~Taylor, M.~V. Sivaselvan and F.~Sachs,
  \emph{Biophysics Journal}, 2023, \textbf{122}, 1--12\relax
\mciteBstWouldAddEndPuncttrue
\mciteSetBstMidEndSepPunct{\mcitedefaultmidpunct}
{\mcitedefaultendpunct}{\mcitedefaultseppunct}\relax
\EndOfBibitem
\bibitem[Ramachandran \emph{et~al.}(2011)Ramachandran, Komura, Seki, and
  Imai]{ram-kom-sek-ima2010}
S.~Ramachandran, S.~Komura, K.~Seki and M.~Imai, \emph{Soft Matter}, 2011,
  \textbf{7}, 1524--1531\relax
\mciteBstWouldAddEndPuncttrue
\mciteSetBstMidEndSepPunct{\mcitedefaultmidpunct}
{\mcitedefaultendpunct}{\mcitedefaultseppunct}\relax
\EndOfBibitem
\bibitem[Young and Stone(2017)]{YoungStone2017_PRF}
Y.-N. Young and H.~A. Stone, \emph{Physical Review Fluids}, 2017, \textbf{2},
  064001\relax
\mciteBstWouldAddEndPuncttrue
\mciteSetBstMidEndSepPunct{\mcitedefaultmidpunct}
{\mcitedefaultendpunct}{\mcitedefaultseppunct}\relax
\EndOfBibitem
\bibitem[Mistriotis \emph{et~al.}(2019)Mistriotis, Wisniewski, Bera, Keys, Li,
  Tuntithavornwat, Law, Perez-Gonzalez, Erdogmus, Zhang, Zhao, Sun, Kalab,
  Lammerding, and
  Konstantopoulos]{mis-wis-ber-key-li-tun-law-per-erd-zha-zha-sun-kal-lam-kon2019}
P.~Mistriotis, E.~O. Wisniewski, K.~Bera, J.~Keys, Y.~Li, S.~Tuntithavornwat,
  R.~A. Law, N.~A. Perez-Gonzalez, E.~Erdogmus, Y.~Zhang, R.~Zhao, S.~X. Sun,
  P.~Kalab, J.~Lammerding and K.~Konstantopoulos, \emph{Journal of Cell
  Biology}, 2019, \textbf{218}, 4093--4111\relax
\mciteBstWouldAddEndPuncttrue
\mciteSetBstMidEndSepPunct{\mcitedefaultmidpunct}
{\mcitedefaultendpunct}{\mcitedefaultseppunct}\relax
\EndOfBibitem
\bibitem[Hou \emph{et~al.}(1994)Hou, Lowengrub, and Shelley]{hou-low-she1994}
T.~Y. Hou, J.~S. Lowengrub and M.~J. Shelley, \emph{Journal of Computational
  Physics}, 1994, \textbf{114}, 312--338\relax
\mciteBstWouldAddEndPuncttrue
\mciteSetBstMidEndSepPunct{\mcitedefaultmidpunct}
{\mcitedefaultendpunct}{\mcitedefaultseppunct}\relax
\EndOfBibitem
\bibitem[Quaife and Biros(2014)]{qua-bir2014}
B.~Quaife and G.~Biros, \emph{Journal of Computational Physics}, 2014,
  \textbf{274}, 245--267\relax
\mciteBstWouldAddEndPuncttrue
\mciteSetBstMidEndSepPunct{\mcitedefaultmidpunct}
{\mcitedefaultendpunct}{\mcitedefaultseppunct}\relax
\EndOfBibitem
\bibitem[Oron \emph{et~al.}(1997)Oron, Davis, and Bankoff]{Oron1997_RMP}
A.~Oron, S.~H. Davis and S.~G. Bankoff, \emph{Rev. Mod. Phys.}, 1997,
  \textbf{69}, 931--980\relax
\mciteBstWouldAddEndPuncttrue
\mciteSetBstMidEndSepPunct{\mcitedefaultmidpunct}
{\mcitedefaultendpunct}{\mcitedefaultseppunct}\relax
\EndOfBibitem
\bibitem[Young \emph{et~al.}(2014)Young, Veerapaneni, and
  Miksis]{Young2014_JFM}
Y.-N. Young, S.~Veerapaneni and M.~J. Miksis, \emph{J. Fluid Mech.}, 2014,
  \textbf{751}, 406--431\relax
\mciteBstWouldAddEndPuncttrue
\mciteSetBstMidEndSepPunct{\mcitedefaultmidpunct}
{\mcitedefaultendpunct}{\mcitedefaultseppunct}\relax
\EndOfBibitem
\end{mcitethebibliography}

\providecommand{\noopsort}[1]{}\providecommand{\singleletter}[1]{#1}%
\providecommand*{\mcitethebibliography}{\thebibliography}
\csname @ifundefined\endcsname{endmcitethebibliography}
{\let\endmcitethebibliography\endthebibliography}{}

\end{document}